\documentclass[%
 reprint,
 amsmath,amssymb,
 aps,
prb
]{revtex4-1}

\usepackage{subfigure}
\usepackage{mathtools}
\usepackage{color}
\usepackage{graphicx}
\usepackage{tabularx}
\usepackage{dcolumn}
\usepackage{bm}

\usepackage[format=hang,justification=raggedright]{caption} 

\newcommand{\Z}{\mathbb{Z}}

\def\p{\partial}

\def\wt{\widetilde}

\newtheorem{thm}{Theorem}[section]

\begin{document}

\preprint{APS/123-QED}

\title{Topological classification under nonmagnetic and magnetic point group symmetry: Application of real-space Atiyah-Hirzebruch spectral sequence to higher-order topology}


\author{Nobuyuki Okuma}\email{okuma@hosi.phys.s.u-tokyo.ac.jp}
\author{Masatoshi Sato}\email{msato@yukawa.kyoto-u.ac.jp}
\author{Ken Shiozaki}\email{ken.shiozaki@yukawa.kyoto-u.ac.jp}
\affiliation{Yukawa Institute for Theoretical Physics, Kyoto University, Kyoto 606-8502, Japan}

%
\date{\today}

\begin{abstract} 
We classify time-reversal breaking (class A) spinful topological crystalline insulators with crystallographic non-magnetic (32 types) and magnetic (58 types) point groups. 
The classification includes all possible magnetic topological crystalline insulators protected by point group symmetry.
Whereas the classification of topological insulators is known to be given by the $K$-theory in the momentum space, computation of the $K$-theory has been a difficult task in the presence of complicated crystallographic symmetry. Here we consider the $K$-homology in the real space for this problem, instead of the $K$-theory in the momentum space, both of which give the same topological classification. We apply the Atiyah-Hirzebruch spectral sequence (AHSS) for computation of the $K$-homology, which is a mathematical tool for generalized (co)homology. 
In the real space picture, the AHSS naturally gives the classification of higher-order topological insulators at the same time. By solving the group extension problem in the AHSS on the basis of physical arguments, we completely determine possible topological phases including higher-order ones for each point group. Relationships among different higher-order topological phases are argued in terms of the AHSS in the $K$-homology. We find that in some nonmagnetic and magnetic point groups, a stack of two $\Z_2$ second-order topological insulators can be smoothly deformed into non-trivial fourth-order topological insulators, which implies non-trivial group extensions in the AHSS.


\end{abstract}

\maketitle

\section{INTRODUCTION}
The topological band theory\cite{hasan,xlq} is one of the major topics in condensed-matter physics.
The earliest example of topological quantum phenomena is the integer quantum Hall effect\cite{thouless}, which is described by the Chern number $n\in\Z$ of occupied bands\cite{kohmoto}. Kane and Mele generalized this idea to two-dimensional insulators with keeping time-reversal symmetry, and found that their intrinsic topological phase is characterized by the $\Z_2$ invariant\cite{kanemele}. 
After this work, further generalization to insulators and superconductors has been done in arbitrary dimensions, which are now called topological insulator and topological superconductor.
The topological classification under time-reversal, particle-hole, and chiral symmetries are summarized in the celebrated topological periodic table \cite{schnyder,kitaev,ryu}.

In addition to these on-site symmetries, topological phases protected by crystalline symmetry have been also explored in insulators\cite{lfu,teofukane} and superconductors\cite{mizushima, teohughes,ueno,zhangkanemele}.
Mathematically, the topological crystalline insulators/superconductors are properly described by the twisted equivariant $K$-theory\cite{freedmoore,thiang,gomi}.
However, computation of the $K$-theory is difficult, and only a limited class of crystalline symmetry has been taken into account so far\cite{chiu,morimoto,cfang,alexandradinata,shiozakisato,cxliurx,cfanglfu,shiozakisatogomi1,shiozakisatogomi2,zwang,shiozakisatogomi3}.

To accomplish topological classification under complicated crystalline space groups, we introduce here a comprehensive mathematical method into condensed matter physics.
In terms of mathematics, the $K$-theory is categorized as a generalized cohomology theory.
The Atiyah-Hirzebruch spectral sequence (AHSS)\cite{AHSS} is known to be a powerful tool to calculate generalized cohomology.
In the AHSS, the space considered is divided into a finite number of simpler cells on which  crystalline symmetry acts as merely on-site-symmetry or relates different cells.
We start from simpler topological classification on the simple cells,  
then check systematically the connectivity of different cells by using the so-called differential maps.
By connecting the cells smoothly, we obtain the desired topological classification on the whole space.
%

The idea of the AHSS has been applied to the $K$-theory in momentum space\cite{shiozaki2}. 
It has been shown that the lowest-order differential map is nothing but the compatibility relation in the band theory, and thus the AHSS naturally fits into the topological band theory.\cite{kruthoff,hcpo,bradlyn,haruki,rxz} 
Moreover,
topological structures beyond the band compatibility relation are obtained via  higher-order differential maps.
A complete list of topological numbers has been obtained for 
non-interacting fermions under 230 space groups without time-reversal and/or particle-hole symmetries\cite{shiozaki2}. 

In addition to the above momentum space picture, there is a real space picture of topological classification\cite{hsong,hcpo2,sjhuang,thorngren,shiozaki1}.
While the momentum space picture is only applicable to non-interacting systems, the real-space picture can be generalized to symmetry protected topological (SPT) phases with many-body interactions.
It  has been suggested that SPT phases are classified in a unified manner by generalized homology in real space.\cite{shiozaki1}
The latter picture can also naturally describe higher-order topological phases\cite{Fang-Fu17,higerti,khalaf,Luka,ono,RasmussenLu18,ElseThorngren18,SongFangQi,ChengWang18,song,calugaru}, which are manifest in lower-dimensional real subspaces.
The AHSS works also in generalized homology as well as the $K$-theory.
In the real space picture of topological insulators, the lowest-order differential map (also called as the boundary map) in the AHSS can be given as the induced representation from higher- to lower-dimensional cells. 

In this paper, we classify topological phases of non-interacting fermions under nonmagnetic and magnetic point group symmetries in terms of the real space picture.
We consider the $K$-homology in real space instead of the $K$-theory in momentum space, both of which give the same topological classification. 
For each point group symmetry, we systematically calculate the $E^\infty$-page in the AHSS for the $K$-homology, which determines topological numbers of higher-order topological insulators. In order to obtain topological numbers on the whole space, we solve the group extension problem in the AHSS by using physical considerations based on Dirac Hamiltonians.
 By combining calculations of the AHSS and such physical considerations, we complete the topological classification under point group symmetries.

This paper is organized as follows.
In Sec. \ref{sec2}, we formulate the AHSS defined in real space as a mathematical tool of topological classification under symmetries.
In the construction of the AHSS, we use the terminology of SPT phases as well as that of topological insulators.
In particular, we demonstrate that the mathematical notion of $E^\infty$-page in the AHSS naturally classifies higher-order topological insulators.
In Sec. \ref{sec3}, we classify topological phases of noninteracting spinful fermions under 32 nonmagnetic point groups, in the absence of time-reversal symmetry.
As an example, we explicitly compute the AHSS under two-fold rotation symmetry in two dimensions.
The classification table is summarized in Fig. \ref{fig4}.
In Sec. \ref{sec4}, we classify topological phases of noninteracting spinful fermions under 58 magnetic point groups.
In contrast with the cases of nonmagnetic point groups, there exist antiunitary operations, which provides an additional complication in the computation of the AHSS.
As an example, we describe the explicit calculation of the AHSS under $2'$ symmetry in three dimensions, which correctly reproduces the $\Z_2$ classification of the second-order topological insulator.
The classification table is summarized in Fig. \ref{fig6}.
In Sec. \ref{sec5}, we explain how to solve the group extension problem in the AHSS.
While the AHSS enables us to obtain the $E^\infty$-page systematically, we need to solve the group extension problem to determine the whole topological structure.
For this purpose, we introduce a Dirac Hamiltonian and consider its adiabatic deformation with adding symmetry-preserving mass terms.
From physical arguments, 
we derive a simple criterion for the non-trivial group extension, by which we solve the group extension problem completely.

\section{Formalism\label{sec2}}
In this section, we introduce the AHSS in real space as a mathematical tool of topological classification. We first describe the role of the AHSS and then formulate it in terms of SPT phases and topological insulators. In particular, we demonstrate that the mathematical notion $E^\infty$-page in the AHSS gives topological classification of higher-order topological insulators.
The following formalism can be understood without the mathematical knowledge of generalized (co)homology. 
See also Refs. [\onlinecite{shiozaki2}] and [\onlinecite{shiozaki1}] for physical interpretations of these mathematical concepts. 

\subsection{Role of AHSS in topological physics}
The key features of topological classification based on the AHSS are summarized as follows:
\begin{itemize}
\item The AHSS is a mathematical tool to compute generalized (co)homology.
\item Classification of SPT phases is done systematically in the framework of generalized (co)homology\cite{kitaev2011,kitaev2,kitaev3,czxiong,czxiong2,gaiotto,shiozaki2,shiozaki1,ElseThorngren18}.
\item For non-interacting topological insulators, the generalized homology reduces to the $K$-homology in real space \cite{shiozaki1,gomigct}.
\end{itemize}
The $K$-theory in momentum space has been used in conventional classification of non-interacting topological insulators.
Owing to the mathematical duality, both of the $K$-theory and $K$-homology give the same classification. 
We here adapt the $K$-homology in real space since it naturally classifies higher-order topological phases at the same time:
The $E^{\infty}$-page defined in the AHSS of generalized homology directly describes higher-order topological phases, as discussed below.
In this paper, we explain how to apply the AHSS to the $K$-homology in real space, providing classification of non-interacting topological insulators.
\begin{figure}[]
\begin{center}
　　　\includegraphics[width=5cm,angle=0,clip]{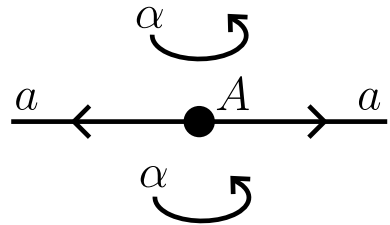}
　　　\caption{Cell decomposition with orientation under two-fold rotation in two dimensions. $A$, $a$, and $\alpha$ represent 0-, 1-, and 2-cells, respectively. The same name is assigned to equivalent cells under the symmetry. $A$ has the on-site symmetry $C_2$, while $a$ and $\alpha$ have no on-site symmetries.}
　　　\label{fig1}
\end{center}
\end{figure}

\subsection{Outline of AHSS}
We here describe the AHSS in terms of SPT phases defined on real space.
Let $X$ be the $d$-dimensional real space, typically taken as the Euclidean space\cite{note1} $\mathbb{E}^d$, and $G$ a symmetry group acting on $X$ such as point and space groups.
SPT phases are classified by a generalized homology 
\begin{align}
h_0^G(X)=\Z^m\bigoplus_{i} \Z_{k_i},
\end{align}
where $m$ and $k_i$ are integers.

The AHSS gives the information of $h_0^G(X)$ in the following way. 
Let us consider a cell decomposition of $X$ that respects symmetry $G$.
We take the cell decomposition so that an element of $G$ acts on each cell as on-site symmetry or it moves the cell to different equivalent cells.
For such a situation,
the following results do not depend on the choice of the cell decomposition. An example of the cell decomposition under two-fold rotation is given in Fig. \ref{fig1}.
Then, we define the $p$-skeleton $X_p$ of $X$ as the set of all cells whose dimensions are equal to or less than $p$\cite{skeleton}:
\begin{align}
X_0\subset X_1\subset \cdots \subset X_d=X.
\end{align}
We can obtain a set of SPT phases on $X$ by embedding $p$-dimensional SPT phases on $X_p$ into the whole space $X$,  which is denoted by $F_p h_0^G(X)$.
Then the following relations hold:
\begin{align}
0\subset F_0h_0^G(X)\subset  F_1h_0^G(X)\subset\cdots \subset F_dh_0^G(X)=h_0^G(X),\notag\\
\end{align}
with
\begin{align}
E^\infty_{p,-p}\simeq F_ph_0^G(X)/ F_{p-1}h_0^G(X),\label{quotient}
\end{align}
or equivalently with the short exact sequence
\begin{align}
0\rightarrow F_{p-1}h_0^G(X)\rightarrow F_{p}h_0^G(X)\rightarrow E^{\infty}_{p,-p}\rightarrow0.
\label{exact}
\end{align}
Here the quotient group in Eq. ($\ref{quotient}$) is called $E^{\infty}$-page.
In the next subsection, we explain how the $E^\infty$-page is obtained in the framework of the AHSS.

Once we obtain the $E^\infty$-page, $F_ph_0^G(X)$ and $h_0^G(X)$ are determined by solving Eq. ($\ref{exact}$).
Mathematically, this problem is known as group extension.
Possible group extensions are not unique in general if $E^{\infty}_{p,-p}$ in Eq. ($\ref{exact}$) contains a torsion subgroup $\bigoplus_i\Z_{k_i}$. In such a case, we need to combine other methods to determine the homology completely.
Eventually, the obtained $h_0^G(X)$ fully classifies SPT phases on $X$.  
 
Although the $E^\infty$-page is introduced as a tool to calculate $h^G_0(X)$,
it has its own physical meaning.
Since $F_ph_0^G$ is obtained from SPT phases on cells whose dimensions are equal to or less than $p$,
$E^\infty_{p,-p}$, the quotient group in Eq. ($\ref{quotient}$), corresponds to SPT phases on the $p$-dimensional submanifold consisting of $p$-cells and their boundaries.
For $p<d$, such embedded topological phases are known as higher-order topological phases.
Therefore, the AHSS in real space naturally classifies both topological insulators and higher-order ones through $h_0^G(X)$ and $E^\infty$-page.
We discuss the relation between them in the last section.

\begin{figure*}[]
\begin{center}
　　　\includegraphics[width=17cm,angle=0,clip]{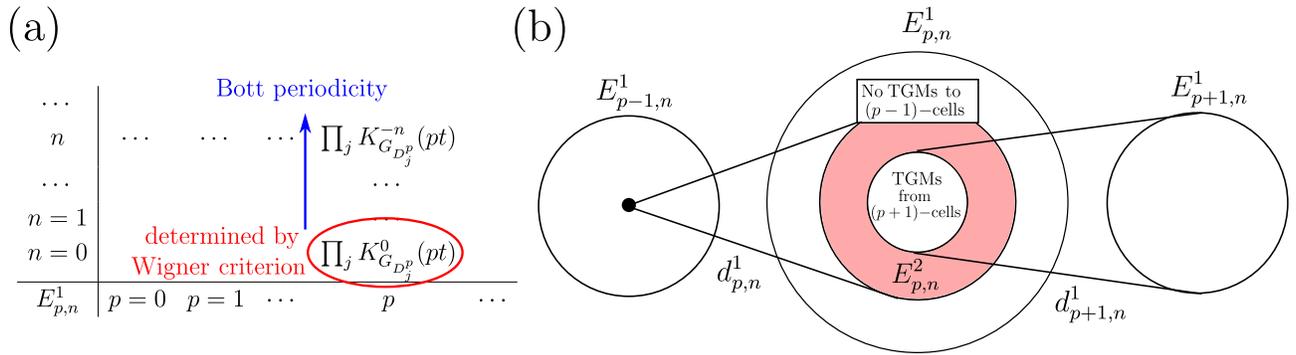}
　　　\caption{(a) Summary of the procedure to calculate the $E^1$-page of the $K$-homology. (b) Schematic picture of the differential map $d^1$. The colored region corresponds to $E^2_{p,n}$ defined in Eq. ($\ref{e2page}$). TGM denotes the topological gapless mode.}
　　　\label{fig2}
\end{center}
\end{figure*}

\subsection{$E^\infty$-page in $K$-homology and higher-order topology}
In the following, we consider the $K$-homology $K_0^G(X)$.
For convenience, we simultaneously treat $K_n^G(X)$, where $n$ is an integer grading.
In the famous topological periodic table, $n$ specifies the complex ($n=0,1$) or real ($n=0,1,\cdots,7$) Altland-Zirnbauer (AZ) classes\cite{altland}.
In general, $K_n^G(X)$ is equipped with additional $n$ chiral symmetries in addition to $G$.
On the cell decomposition of the AHSS, $G$ reduces to on-site symmetry on each cell, and thus we can specify the corresponding AZ class. The additional $n$ chiral symmetries shift the AZ class accordingly.
As in the case of $n=0$, there exists a short exact sequence
\begin{align}
0\rightarrow F_{p-1}K_n^G(X)\rightarrow F_{p}K_n^G(X)\rightarrow E^\infty_{p,n-p}\rightarrow0.
\end{align}

Now we would like to explain how to obtain the $E^\infty$-page.
As the first step,  we introduce the $E^1$-page as follows.
Let $D_j^p$ be a $p$-dimensional cell in the cell decomposition, where $j$ runs the set of inequivalent $p$-cells under $G$.
Then, the $E^1$-page is defined as
\begin{align}
E^1_{p,n}&\equiv \bigoplus_j K^{G_{D_j^p}}_{p+n} (D^p_j,\partial D^p_j),
\end{align}
where $G_{D_j^p}$ is the little group of $G$ on $D_j^p$ (namely, on-site symmetry on $D_j^p$).
Here  $K^{G_{D_j^p}}_{p+n}(D^p_j, \partial D_j^p)$ denotes the relative $K$-homology between $D_j^p$ and its boundary $\partial D_j^p$, which corresponds to  $(p+n)$-th graded SPT phases on the $p$-sphere $D_j^p/\partial D_j^p$ with on-site symmetry $G_{D_j^p}$.
For the non-interacting case, the SPT phases are given by the $K$-theory, so we have\cite{shiozaki1}
\begin{align}
E^1_{p,n}=\bigoplus_j K^{-n}_{G_{D_j^p}}(pt), 
\label{eq:e1_pt}
\end{align}
where $K^{-n}_{G_{D_j^p}}(pt)$ is the $n$-th graded $K$-group of a point with on-site symmetry $G_{D_j^p}$.
Here we have used the Poincar\'{e} duality between the $K$-homology  and the $K$-theory.
We then assign the emergent AZ class on $D_j^p$ by applying the Winger criterion\cite{shiozaki2} on $G_{D_j^p}$, and identify an element of the $K$-group of a point as a set of irreducible representations of $G_{D_j^p}$.
Owing to the Bott periodicity, we can obtain all terms of the $E^1$-page just by calculating the $n=0$ case.

By definition, $E^1_{p,n}$ represents ($p+n$)-th graded SPT phases on the $p$-spheres $D_j^p/\partial D_j^p$, but it is not the only physical meaning.
$E^1_{p,n}$ is also interpreted as ($p+n+1$)-th graded topological gapless modes:
The second interpretation comes from the bulk-boundary correspondence, which claims that anomalous gapless boundary modes exist if the bulk topological phase is nontrivial. 
Actually, in the non-interacting case, we can increase the dimension of the system by 1 with keeping the bulk topological numbers, by shifting the grading as $p+n\to p+n+1$\cite{shiozakisatogomi3}.
Then, we can obtain anomalous gapless modes on $p$-spheres as boundary modes of the $(p+1)$-dimensional system, which are also characterized by $E^1_{p,n}$.

Based on the above consideration, we can define the first differential map (or also called as the boundary map)
\begin{align}
d^1_{p,n}: E^1_{p,n}\rightarrow E^1_{p-1,n},
\label{eq:d1_map}
\end{align}
which relates the $(p+n)$-th graded SPT phases 
on $p$-cells ($E^1_{p,n}$) to the $(p+n)$-th graded anomalous gapless modes 
on the adjacent $(p-1)$ cells ($E^1_{p-1,n}$) by the bulk-boundary correspondence:
If one puts a SPT on a $p$-cell, one obtains its boundary gapless modes on the adjacent $(p-1)$-cells.
In practical terms, $d^1_{p,n}$ is expressed as an integer matrix.
The differentials satisfy the following relation:
\begin{align}
d^1_{p,n}\circ d^1_{p+1,n}=0,\label{boundarymap}
\end{align}
which corresponds to the fact that the boundary of the boundary is nothing.
Owing to Eq. ($\ref{boundarymap}$), we can define the homology of $d^1$ called $E^2$-page:
\begin{align}
E^2_{p,n}\coloneqq  \mathrm{Ker}\ (d^1_{p,n})/\mathrm{Im}\ (d^1_{p+1,n}).\label{e2page}
\end{align}
Physically, $E^{2}_{p,n}$ represents a family of  SPT phases on $p$-cells that do not create anomalous boundary modes on their adjacent $(p-1)$-cells.
Therefore, we can connect $p$-cells and ($p-1$)-cells without creating gapless modes on ($p-1$)-cells.
At the same time, 
by regarding the page as gapless modes\cite{adiabaticpump}, we also exclude anomalous gapless modes from $(p+1)$-cells in $E^2_{p,n}$.
See Fig.\ref{fig2} (b).

In general, this is not the end of the story because the connectivities between $p$-cells and ($p\pm2,3,\cdots$)-cells are not considered in the $E^2$ page. The higher differential and $E^r$-page are iteratively given as 
\begin{align}
&d^{r-1}_{p,n}: E^{r-1}_{p,n}\rightarrow E^{r-1}_{p-r+1,n+r-2},\notag\\
&E^r_{p,n}\coloneqq  \mathrm{Ker}\ (d^{r-1}_{p,n})/\mathrm{Im}\ (d^{r-1}_{p+r-1,n-r+2}).
\end{align}
The $E^r$-page converges at $r=d+1$, and the converged page is called the limiting page $E^{\infty}$.
Note that the $E^r$-page can converge at lower $r$ when $d^s=0$ $(s\ge r)$. 
In  class A with nonmagnetic and magnetic point groups, the $E^2$-page becomes the limiting page, as discussed in the following sections.

We can interpret $E^{\infty}_{p,-p}$ as SPT phases on the $G$-invariant $p$-dimensional submanifold in the original symmetry class with $p+n=0$. In recent terminology, such topological states defined on the lower-dimensional subspace of $X$ are called  higher-order topological insulators. Thus, $E^\infty_{p,-p}$ for $p<d$ classifies $(d-p+1)$-th-order topological insulators, while $E^\infty_{p,-p}$ for $p=d$ classifies conventional topological insulators.

\section{Topological classification under nonmagnetic point group symmetries\label{sec3}}
In this section, we classify topological phases in class A ($n=0$) under non-magnetic point group symmetries acting on spinful fermions.
In class A, time-reversal, particle-hole, and chiral symmetries are absent, so the emergent AZ class for $n=0$ on each cell is trivially A.
In that case, the emergent AZ class for $n=0$ on each cell is trivially A.
Thus, the $K$-group of a point takes the form of $\Z^m$, which is generated by $m$ unitary irreducible representations (irreps) of $G_{D_j^p}$. In other words, the $\Z$ topological number counts the number of states in each irrep.
By taking the sum over inequivalent $p$-cells and using the Bott periodicity for complex AZ classes, we obtain 
\begin{align}
&E^1_{p,2l}=\bigoplus_j \Z^{m_j}\notag\\
&E^1_{p,2l+1}=0,\label{nonmage1}\notag\\
\end{align}
where $l$ is an integer, and $m_j$ is the number of possible unitary irreps of $G_{D_j^p}$.

\begin{figure}[]
\begin{center}
　　　\includegraphics[width=8cm,angle=0,clip]{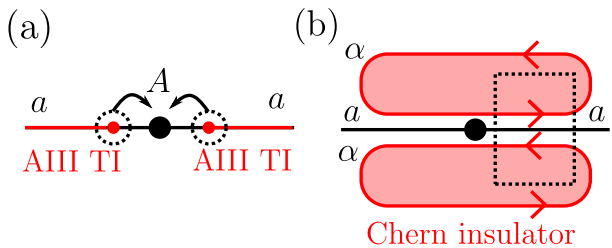}
　　　\caption{Schematic pictures of the first differentials (a) $d^1_{1,0}$ and (b) $d^1_{2,0}$ under the two-fold rotation symmetry.}
　　　\label{fig3}
\end{center}
\end{figure}

In the following, we calculate the first differential $d^1$ and $E^2$-page for the $E^1$-page ($\ref{nonmage1}$).
As an example, we consider two-fold rotation in two dimensions.
We take the cell decomposition defined in Fig. \ref{fig1}.
For each $p$, the number of inequivalent $p$-cells is one.
The 0-cell $A$ has on-site point group symmetry $C_2$, while the 1-cell $a$ and 2-cell $\alpha$ have no symmetry.
There are two irreps under $C_2$ rotation, which are characterized by the rotation eigenvalues $C_2=\pm i$, while there is only one irrep under no symmetry. Thus, the $E^1$-page is given by
\begin{align}
\begin{array}{c|ccc}
n=1 & 0 & 0 & 0 \\
n=0 &  \Z^2& \Z& \Z \\
\hline 
E^1_{p,n} & p=0 & p=1 & p=2 \\ 
\end{array}\ \ \ .
\end{align}
Next, we consider the first differential $d^1_{1,0}$ from $E^1_{1,0}$ to $E^1_{0,0}$.
By definition, $E^1_{1,0}$ describes the 1st graded topological phases on the 1-cell $a$, which corresponds to one-dimensional topological insulators in class AIII. Let us put the nontrivial topological insulator on $a$. 
The one-dimensional topological insulator creates gapless edge states characterized by its topological number $\Z$, which defines the first differential map $d_{1,0}^1$ [Fig. \ref{fig3} (a)]. 
Since the map respects $C_2$ symmetry, two boundary modes are created on $A$ from two $a$s adjacent to $A$, which form irreps at $A$ as induced representations.
%
In other words, the map is nothing but the construction of induced representation\cite{shiozaki1} from $a$ with no symmetry to $A$ with $C_2$ symmetry.
Thus the differential is given by the compatibility relation
\begin{align}
d^1_{1,0}=
\begin{pmatrix}
1 \\
1 \\
\end{pmatrix},\label{d110}
\end{align} 
where rows and  columns represent the irreps on $A$ and $a$, respectively.
This means that a pair of $C_2=\pm i$ irreps are induced on $A$ from a trivial irrep on $a$.

\begin{figure*}[]
\begin{center}
　　　\includegraphics[width=16cm,angle=0,clip]{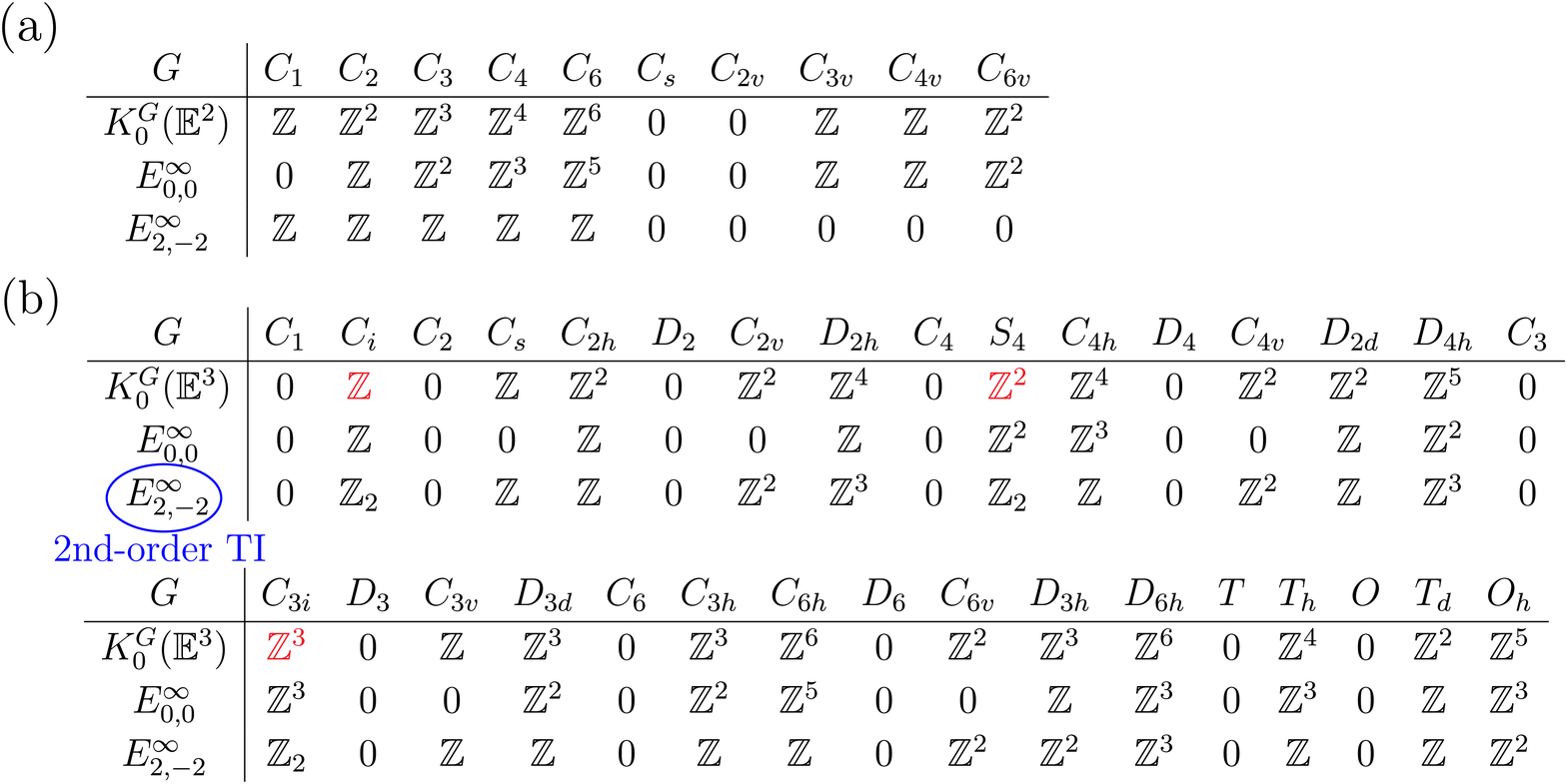}
　　　\caption{Topological classification of complex A class under (a) two- and (b) three-dimensional nonmagnetic point group symmetries for spinful fermions. The red characters represent the nontrivial group extensions. In three dimensions, $E^{\infty}_{2,-2}$ gives the topological classification of the second-order topological insulators (TIs). See Appendix \ref{sec:app_c} for the explicit configurations of the surface edge modes and bound states.}
　　　\label{fig4}
\end{center}
\end{figure*}

The first differential $d^1_{2,0}$ from $E^1_{2,0}$ to $E^1_{1,0}$ can also be determined in the same way. See Fig.\ref{fig3}:
$E^1_{2,0}$ describes a 2nd graded, or equivalently 0th graded topological insulator on the 2-cell $\alpha$ which is nothing but the Chern insulator. On the other hand, $E^1_{1,0}$ represents 0th graded anomalous gapless modes on the 1-cell $a$, {\it i.e.} the chiral anomaly due to chiral edge modes.
In this case, there are no on-site-symmetry for both the 1- and 2-cells.
The Chern insulator on $\alpha$ can create a chiral edge mode, but two $\alpha$s are next to one $a$, and they give edge modes with opposite chirality. 
The contributions from them cancel each other, and thus the differential becomes trivial,
\begin{align}
d^1_{2,0}=0.\label{d120}
\end{align}
This cancellation can be also interpreted in terms of the orientation-relation between the 1-cell $a$ and 2-cell $\alpha$.
It is convenient to assign the sign of orientation to each cell with respecting the symmetry [Fig. \ref{fig1}].
The relative orientation sign between $a$ and $\alpha$ corresponds to the sign of contribution discussed above.
Note that Eqs. ($\ref{d110}$) and ($\ref{d120}$) apparently satisfy Eq. ($\ref{boundarymap}$)

We are now in a position to calculate the $E^2$-page by applying Eq. ($\ref{e2page}$) to Eqs. ($\ref{d110}$) and ($\ref{d120}$).
The result is given by
\begin{align}
\begin{array}{c|ccc}
n=1 & 0 & 0 & 0 \\
n=0 &  \Z& 0& \Z \\
\hline 
E^2_{p,n} & p=0 & p=1 & p=2 \\ 
\end{array}\ \ \ .\label{2rote2page}
\end{align}
Here $E^2_{0,0}=\Z$ is given by $n^{-i}-n^{+i}$, where $n^{\pm i}$ are the number of states in irreps $C_2=\pm i$.
$E^2_{2,0}=E^2_{2,-2}=\Z$ is the Chern number of the Chern insulator on 2-cells.
Since $E^1_{p-1,n}$,  $E^1_{p,n}$, and  $E^1_{p,n+1}$ have no torsion, Eq. ($\ref{e2page}$) can be systematically calculated by using the Smith decomposition.

In this example, the $E^2$-page is nothing but the $E^\infty$-page because of the absence of the higher differentials.
Thus the remaining issue is the group extension problem.
Substituting Eq. ($\ref{2rote2page}$) into Eq. ($\ref{exact}$), we obtain
\begin{align}
&F_0K^G_0(X)=E^\infty_{0,0}=\Z,\\
&F_1K^G_0=E^\infty_{0,0}=\Z,\\
&0\rightarrow \Z\rightarrow K^G_0=F_2K^G_0\rightarrow E^{\infty}_{2,-2}=\Z\rightarrow 0.
\end{align}
This short exact sequence splits since the right side has no torsion, and thus the solution is given by the sum of the left and right sides.
Namely, the $K$-homology is given by
\begin{align}
K^G_0(X)=\Z^2.
\end{align}

The above discussions are simply generalized to any nonmagnetic point group symmetries in two and three dimensions.
Fortunately, the higher differentials for them are 0 even in the case of three dimensions, and thus the $E^2$-pages are always equal to the $E^\infty$-pages.
The obtained results are summarized in Fig.\ref{fig4}.
Here we omit $E^\infty_{1,-1}$ and $E^\infty_{3,-3}$ since they are trivial. (This is because $E^1_{1,-1}$ and $E^1_{3,-3}$ are 0.)
In three dimensions, $E^\infty_{2,-2}$ classifies the second-order topological insulators, whose boundaries host one-dimensional gapless modes. 
Note that the group extension problems for the $C_i$ (inversion), $C_{3i}$, and $S_4$ symmetries cannot be solved without further information because $E^\infty_{2,-2}$ has a torsion.
For example, the short exact sequence for the $C_i$ symmetry is given by
\begin{align}
0\rightarrow\Z\rightarrow K^G_0(X)\rightarrow \Z_2\rightarrow 0.\label{invextention}
\end{align}
The trivial solution of Eq. ($\ref{invextention}$) is  $K_0^G(X)=\Z\oplus\Z_2$, but 
$K_0^G(X)=\Z$ is also a solution of Eq. ($\ref{invextention}$).
In general, if the group extension problem has a non-trivial solution, then  we need an additional method other than the AHSS to fully determine the $K$-homology.
In the above case of $C_i$ symmetry, the true answer is $K_0^G(X)=\Z$. Using physical arguments, we solve the group extension problem in the last section.

\section{Topological classification under magnetic point group symmetries\label{sec4}}
In this section, we classify topological phases of complex A class ($n=0$) under magnetic point group symmetries.
In general, a magnetic point group $G'$ is related with a nonmagnetic point group $G$:
\begin{align}
G'&=G+a_0G,\\
a_0&=Tv_0,
\end{align}
where $v_0$ is a symmetry operation that is not an element of $G$, and $T$ is the time-reversal operation.
Note that this is a narrow definition of magnetic point group symmetry in which $v_0$ is not the identity operation. There exist 58 types of such magnetic point groups.
Very recently, Song $et\ al.$ have performed topological classification under pure time-reversal and nonmagnetic point group symmetries except for the part corresponding to $E^\infty_{0,0}$ \cite{song}. We here consider the cases with $v_0\neq1$.

To determine the $E^1$-page, we need irreps of $G'$.
Following the standard recipe\cite{bradley}, 
we start from irreps of $G$.
Then, the Wigner criterion detects how the irreps of $G$ behave under $a_0$.
Let us consider an irrep $\alpha$ of $G$.
In the Wigner criterion, we use the following quantity:
\begin{align}
W_{\alpha}=\frac{1}{|G|}\sum_{g\in G}z_{a_0g,a_0g}\chi_{\alpha}((a_0g)^2),
\end{align}
where $\chi_{\alpha}$ is the character of the irrep $\alpha$.
The factor system of $G'$, $\{z_{g,h}=\pm1|g,h\in G'\}$, is defined as 
\begin{align}
z_{g,h}U(gh)=U(g)U(h),
\end{align}
where 
\begin{align}
U(g)=
\begin{cases}
\mathcal{U}(g)&\ \mathrm{for}\ g\in G\\
\mathcal{U}(g)K&\ \mathrm{for}\ g\in a_0G
\end{cases}
\end{align}
is a projective representation of $G'$ with $\mathcal{U}(g)$ and $K$ being a unitary matrix and the conjugate operator, respectively.

$W_\alpha$ takes the values $\pm1,0$ and gives the information of the degeneracy generated by $a_0$:
\begin{align}
\begin{array}{rc}
W_\alpha& \mbox{Degeneracy of irrep $\alpha$}  \\
\hline 
\hline 
1&  \mbox{No additional degeneracy}\\
\hline
-1 & \mbox{Kramers degeneracy}\\ 
\hline
0 & \mbox{$\alpha$ has the conjugate irrep $\bar{\alpha}$ of $G$}  \\ 
& \mbox{($\alpha$ and $\bar{\alpha}$ are interchanged by $a_0$)}\\
\hline 
\end{array}\ \ \ .\notag
\end{align}
The results of the Wigner criterion for 58 magnetic point groups are listed in Ref. [\onlinecite{bradley}].
$W_\alpha$ determines the emergent AZ class, and thus it determines the $K$-group of a point for irrep $\alpha$, 
\begin{align}
\begin{array}{rcc}
W_\alpha&\mathrm{Emergent\ AZ\ class}&\mbox{$K$-group}\\
\hline 
\hline 
1& \mathrm{AI} &\Z\\
-1 &\mathrm{AII} & \Z \\ 
0 & \mathrm{A} &\Z \\
\hline  
\end{array}\ \ \ .\notag
\end{align}
Using this rule, we can determine $E^1_{p,0}$ as in the case of nonmagnetic point groups.

\begin{figure}[]
\begin{center}
　　　\includegraphics[width=5cm,angle=0,clip]{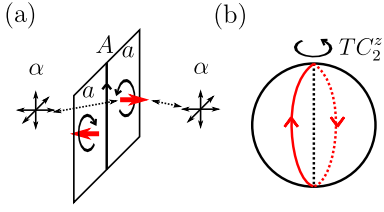}
　　　\caption{(a) Cell decomposition with orientation for magnetic point group $2'$ in three dimensions. The left $\alpha$ has the same orientation as that of $a$, while the other one s the opposite orientation. (b) Surface edge mode of second-order topological insulator under $2'$ symmetry. A gapless mode appears at the edge of the Chern insulator on the 2-cells.}
　　　\label{fig5}
\end{center}
\end{figure}

In the case of magnetic point groups, the first differential can be defined as a map between different emergent AZ classes.
To see this, we consider the magnetic point group $2'$ acting on spinful fermions in three dimensions.
We take the cell decomposition defined in Fig. \ref{fig5}(a).
The 1-cell $A$ has on-site magnetic symmetry
\begin{align}
2'=C_1+(TC^z_2 )C_1,
\end{align}
where $C^z_2$ is two-fold rotation around the $z$ axis, and $C_1$ denotes the trivial group consisting of only the identity operation $E$.
The irrep of $C_1$ is the  one-dimensional representation counting the number of states.
According to the Wigner criterion, the emergent AZ class on the 1-cell $A$ is AI:
\begin{align}
W=\chi((TC^z_2)^2)=1.
\end{align}
The 2-cell $a$ and 3-cell $\alpha$ have no on-site symmetry.
In summary, the $E^1$-page is given by
\begin{align}
\begin{array}{c|cccc}
n=7 & 0 &0 &0&0\\
n=6 & 0 &0 &\Z&\Z\\
n=5 & 0 & 0& 0&0 \\
n=4 & 0 &\Z &\Z&\Z\\
n=3 & 0 & 0 & 0&0 \\
n=2 & 0& \Z_2& \Z&\Z \\
n=1 & 0 & \Z_2 & 0&0 \\
n=0 & 0& \Z& \Z&\Z \\
\hline 
E^1_{p,n} & p=0 & p=1 & p=2 &p=3\\ 
\end{array}\ \ \ .
\end{align}
To determine the $E^1$-page on the 1-cell ($p=1$) in the above, we have used the Bott periodicity for real AZ classes
\begin{align}
\Z\rightarrow\Z_2\rightarrow\Z_2\rightarrow0\rightarrow\Z\rightarrow0\rightarrow0\rightarrow0.
\end{align}

\begin{figure*}[]
\begin{center}
　　　\includegraphics[width=16cm,angle=0,clip]{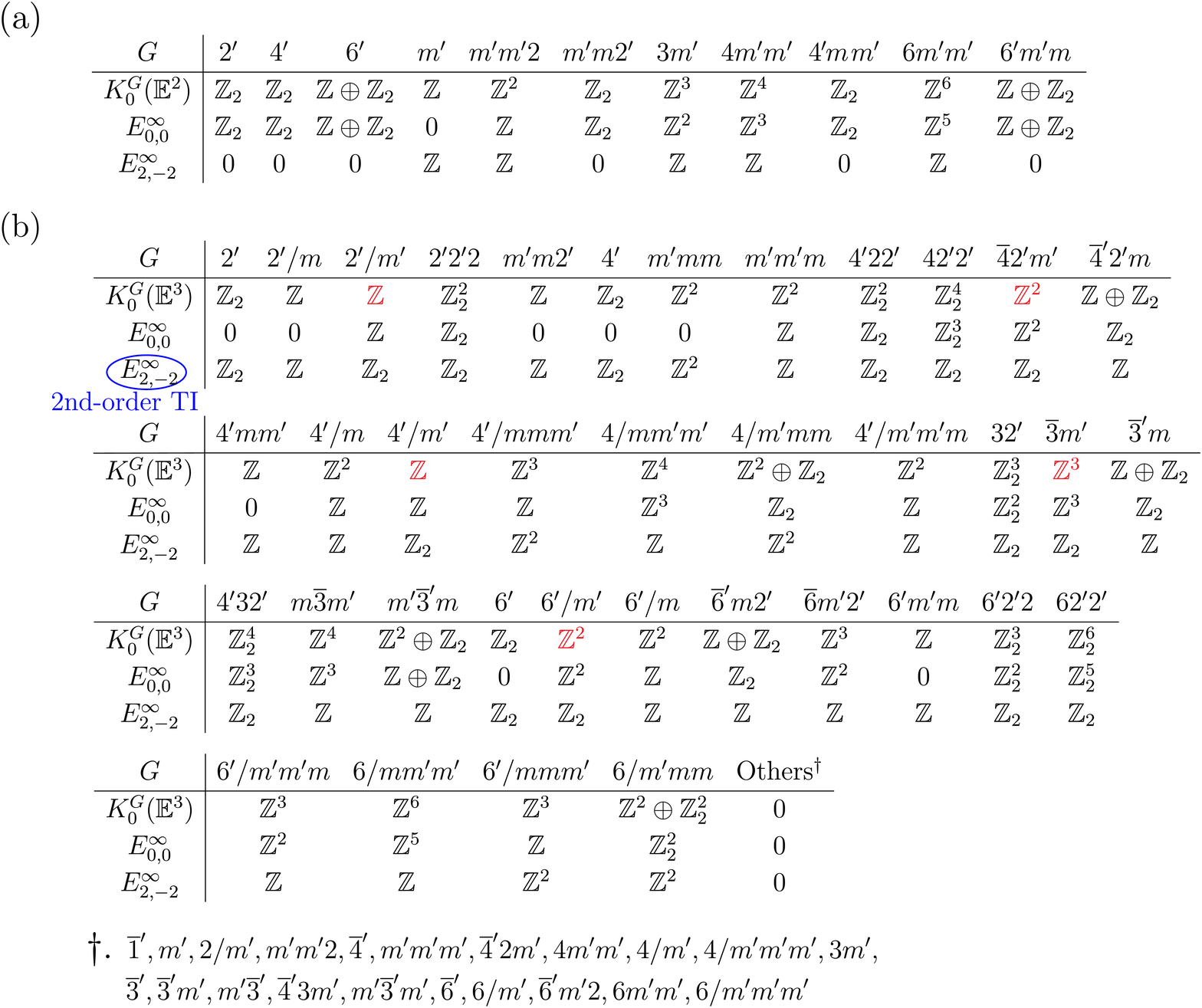}
　　　\caption{Topological classification of complex A class under (a) two- and (b) three-dimensional magnetic point group symmetries for spinful fermions. The red characters represent the nontrivial group extensions. In three dimensions, $E^{\infty}_{2,-2}$ gives the topological classification of the second-order topological insulators (TIs). See Appendix \ref{sec:app_c} for the explicit configurations of the surface edge modes and bound states.}
　　　\label{fig6}
\end{center}
\end{figure*}

As in the case of nonmagnetic point groups, we use the compatibility relation to obtain the first differentials.
However, two additional points should be considered in the case of magnetic point groups.
First, we need the compatibility relation between different AZ classes.
The rule for compatibility relation is summarized in the Appendix.
In any case, we can construct the compatibility relation from that of the nonmagnetic parts.
For instance, $d^1_{2,0}$ is the differential map from class A to class AI.
The compatibility relation in this case is given by doubling the nonmagnetic part.
In this simple case, the differential for the nonmagnetic part is just the identity and thus we obtain
\begin{align}
d^{1}_{2,0}=2.
\end{align}
Second, the shift of the grade can change the original time-reversal-type magnetic symmetry into the particle-hole-type one,
\begin{align}
a_0'Ha_0'^{-1}=-H,
\end{align}
where $a'_0$ is the antiunitary operator in the shifted grade, which is constructed  from $a_0$.
In our problem, this change occurs in the cases of $n=2,3,6,7$.
In particular, the change for $n=6$ is crucial because $E^\infty_{2,6}=E^{\infty}_{2,-2}$ enters into the short exact sequence ($\ref{exact}$).
Let us calculate $d^1_{3,0}$ and $d^1_{3,6}$.
The differential from the 3-cell $\alpha$ to the 2-cell $a$ is determined by assigning the sign of orientation relation between them.
For $n=0$, contributions from two $\alpha$s to $A$ cancel each other out.
On the other hand, for $n=6$, there arises a  sign change coming from the 
fact that particle-hole-like symmetry changes a state into the anti-state.
As a result, we can obtain a non-zero $d^1_{3,6}$.
In summary, we have
\begin{align}
d^1_{3,0}=0,\quad
d^1_{3,6}=2.
\end{align}

Other first differentials can be obtained in a similar manner, by which we calculate the $E^2$-page.
The result of the $E^2$-page for $2'$ is 
\begin{align}
\begin{array}{c|ccccc}
n=7 &0& 0 & 0 & 0 \\
n=6 &0& 0 & \Z_2 & 0 \\
n=5 &0& 0 & 0 & 0 \\
n=4 &0& 0 & 0 & \Z \\
n=3 &0& 0 & 0 & 0 \\
n=2 &0& 0 & 0 & 0 \\
n=1 &0& \Z_2 & 0 & 0 \\
n=0 &0& \Z_2 & 0& \Z \\
\hline 
E^2_{p,n} & p=0 & p=1 & p=2& p=3
\end{array}.
\end{align}
While the second differential from $E^2_{3,0}$ to $E^2_{1,1}$ can be nontrivial, there are no higher differentials acting on $E^2_{p,-p}$.
Thus, $E^2_{p,-p}=E^\infty_{p,-p}$, and the topological classification is given by
\begin{align}
&E^{\infty}_{0,0}=E^{\infty}_{1,-1}=E^{\infty}_{3,-3}=0,\\
&E^{\infty}_{2,-2}=\Z_2,\\
&K^G_0(\mathbb{E}^3)=\Z_2.
\end{align}
The $\Z_2$ topological phase in the above is  a second-order topological insulator \cite{benjamin,bjyang} [Fig.\ref{fig5}(b)].
The generator of $\Z_2$ is Chern insulators on 2-cells that respect $2'$ symmetry.

Using the method in the above, we have calculated the $E^2$-pages for all 58 magnetic point groups and found that there are no higher differentials that affect  $E^2_{p,-p}$.
Therefore, we have  $E^2_{p,-p}=E^{\infty}_{p,-p}$, as in the case of nonmagnetic point groups.
The results are summarized in Fig. $\ref{fig6}$.
Note that the triviality of third differentials from $E^3_{3,-2}$ to $E^3_{0,0}$ is checked only after the explicit calculations.

\section{Physics of group extension\label{sec5}}
Finally, we describe the nontrivial group extension by using the language of topological physics.
We again consider the symmetry $C_i$ (inversion).
As mentioned above, the group extension is nontrivial:
\begin{align}
0\rightarrow E^\infty_{0,0}=\Z\rightarrow K^G_{0}(\mathbb{E}^3)=\Z\rightarrow E^\infty_{2,-2}=\Z_2\rightarrow0.
\end{align}
Here the generator of $E^\infty_{2,-2}=\Z_2$ is a Chern insulator that respects the inversion symmetry. 
Since the classification of the second-order topological insulator $E^\infty_{2,-2}$ is $\Z_2$, a stack of two nontrivial states is trivial in $E^\infty_{2,-2}$, and thus two surface edge modes can be gapped out.
When the group extension is nontrivial, however, such a trivial state in $E^\infty_{2,-2}$ is a nontrivial state in $E^\infty_{0,0}$.
This can be understood by using adiabatic transformations, in which the gap is not closed.

\begin{figure}[]
\begin{center}
　　　\includegraphics[width=7cm,angle=0,clip]{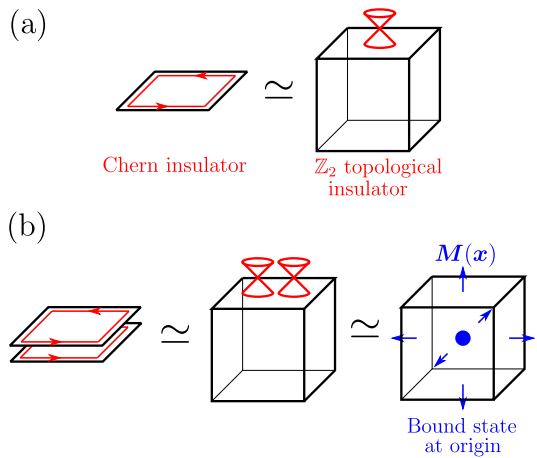}
　　　\caption{Schematic pictures of adiabatic transformations under the inversion symmetry $C_i$ and no time-reversal symmetry. (a) Equivalence between the Chern insulator with the Chern number: $+1$ and a $\Z_2$ topological insulator. (b) Adiabatic process representing nontrivial group extension. A stack state of two Chern insulators can be deformed into one bound state at the origin surrounded by the hedgehog-like vector field $\bm{M}(\bm{x})$, which corresponds to the generator of $E^\infty_{0,0}=\mathbb{Z}$. }
　　　\label{fig7}
\end{center}
\end{figure}

Let us consider the Chern insulator with the Chern number: +1 respecting the inversion symmetry, which is a generator of $E^\infty_{2,-2}=\Z_2$.
Without changing the symmetry, this state can be adiabatically transformed into the three-dimensional $\Z_2$ topological insulator described by a $4\times4$ Dirac Hamiltonian (see Appendix B for details) [Fig.\ref{fig7}(a)]:
\begin{align}
H_{3d}&=-i\sum_i \partial_i\sigma_i\tau_x+m\tau_z,\notag\\
I_{3d}&=\tau_zP(\bm{x}\rightarrow-\bm{x}),
\end{align} 
where $m$ is the mass, $\sigma$s and $\tau$s represent the $2\times2$ Pauli matrices in spin and parity spaces, respectively, and $P$ is the operator that changes $\bm{x}$ to $-\bm{x}$.
The inversion operator $I_{3d}$ does not change the Hamiltonian $H_{3d}$. 
Instead of the Chern insulators, we consider a stack of two topological insulators:
\begin{align}
H_{3d}\oplus H_{3d}&=-i\sum_i \partial_i\sigma_i\tau_x\mu_0+m\tau_z\mu_0,\notag\\\label{stackham}
I_{3d}\oplus I_{3d}&=\tau_z\mu_0 P,
\end{align} 
where $\mu$ is an identity matrix in the stack space.
Under the inversion symmetry, the following additional mass term can exist:
\begin{align}
&\bm{M}(\bm{x})\cdot\bm{\mu}\tau_y,
\end{align}
where the vector field $\bm{M}$ obeys the constraint from the inversion symmetry:
\begin{align}
\bm{M}(-\bm{x})=-\bm{M}(\bm{x}).\label{constraint}
\end{align}
This mass term can be added to the Hamiltonian ($\ref{stackham}$) without the gap closing.
It is known that the number of bound states at the origin of such a Hamiltonian is equal to the winding number of the vector field $\bm{M}$ surrounding the origin, which is a consequence of the index theorem \cite{callias,weinberg,niemi,teokane,volovik}.
Under the constraint ($\ref{constraint}$), the winding number can only be an odd number, which means that the Hamiltonian ($\ref{stackham}$) cannot be adiabatically transformed into the state without the bound state.
Such bound states at the origin are nothing but elements of $E^\infty_{0,0}$.
By taking the winding number to be 1, the stack of two three-dimensional topological insulators, or equivalently that of two second-order topological insulators, can be transformed into the state that corresponds to the generator of $E^\infty_{0,0}$ [Fig.\ref{fig7}(b)].

The above discussion can be simply generalized to the other nonmagnetic and magnetic point groups with $E^{\infty}_{2,-2}=\mathbb{Z}_2$ (see Appendix B for details).
We focus on the behavior of the mass term $\bm{M}$ under the inversion, rotation, and time-reversal operations since all symmetric operations of point groups are combinations of them.
Each symmetric operation acts on the mass term  in the following ways (see Appendix for details):
\begin{align}
\mathrm{Inversion}&:\ \bm{M}(\bm{x})\rightarrow-\bm{M}(-\bm{x}),\notag\\
\mathrm{Rotation}&:\  \bm{M}(\bm{x})\rightarrow \bm{M}(\hat{R}_{\bm{n},\theta}\ \bm{x}),\notag\\
\mathrm{Time\ reversal}&:\  \bm{M}(\bm{x})\rightarrow (-M_1(\bm{x}),M_2(\bm{x}),-M_3(\bm{x})),
\label{eq:condition_m_vector}
\end{align}
where $\hat{R}_{\bm{n},\theta}$ is a rotation matrix.
The group extension is nontrivial if and only if $\bm{M}$ can be taken as a uniform vector field (see Appendix for details), whose winging number is equal to zero, under the above conditions.
By using this property, we determine the $K$-homology for the cases with $E^{\infty}_{2,-2}=\mathbb{Z}_2$ [Figs. \ref{fig4} and \ref{fig6}].
There are two types of exact sequences that can not be determined only by the AHSS:
\begin{align}
&0\rightarrow \Z_2^p\rightarrow K^G_0(\mathbb{E}^3)\rightarrow \Z_2\rightarrow0,\label{former}\\
&0\rightarrow \Z^q\rightarrow K^G_0(\mathbb{E}^3)\rightarrow \Z_2\rightarrow0,\label{latter}
\end{align}
where $p,q$ are integers.
In the cases described by Eq. ($\ref{former}$), we have checked that there can exist a uniform vector field.
Thus, the group extension is trivial:
\begin{align}
K^G_0(\mathbb{E}^3)=\Z_2^{p+1}.
\end{align}
As an Abelian group, the solution of the short exact sequence ($\ref{latter}$) can have the following two forms:
\begin{align}
K^G_0(\mathbb{E}^3)&=\Z^q+\Z_2,\\
K^G_0(\mathbb{E}^3)&=\Z^q.\label{latter2}
\end{align}
We have checked that there cannot exist a uniform vector field under the point groups described by Eq. ($\ref{latter}$).
Thus, the group extension is nontrivial, and the true solution is Eq. ($\ref{latter2}$).

\section{Summary}
We have classified the (class A) topological phases of noninteracting spinful fermions under nonmagnetic and magnetic point groups.
We have considered the $K$-homology of real space instead of the $K$-theory of momentum space, both of which give the same topological classification, and computed it in the framework of the Atiyah-Hirzebruch spectral sequence (AHSS). In the real space picture, the mathematical notion $E^\infty$-page introduced in the AHSS  naturally gives the classification of the higher-order topological insulators. We have systematically determined the $E^\infty$-page and derived the short exact sequence that contains the $K$-homology for each point group. Mathematically, the $K$-homology is given as a solution of the group extension problem.
The consideration of the relationship between the $E^\infty$-page and $K$-homology plays an important role to solve the group extension problem.
We have found that in some nonmagnetic and magnetic point groups, a stack of two $\Z_2$ second-order topological insulators can be smoothly deformed into a nontrivial fourth-order topological insulator, which implies a nontrivial group extension in the AHSS.
For each point group, we have determined the $K$-homology in addition to the $E^\infty$-page and summarized them in tables.

\begin{acknowledgments}
This work was supported by a Grant-in-Aid for Scientific Research on Innovative Areas "Topological Materials Science" (KAKENHI Grant No. JP15H05855) from the Japan Society
for the Promotion of Science (JSPS). M. S. was supported by KAKENHI Grant No. JP17H02922 from the JSPS.
N. O. was supported by KAKENHI Grant No.18J01610.
K. S. was supported by PRESTO, JST (JPMJPR18L4).
\end{acknowledgments}

\appendix

\section{Rules for compatibility relation}
In this appendix, we summarize the compatibility relation between a ($p-1$)-cell and adjacent $p$-cells in the cases with antiunitary symmetries.
We write down the rules for each emergent AZ class at $n=0$.
$i$ and $\lambda$ denote representations of the unitary part of the on-site symmetries at ($p-1$)- and $p$-cells, respectively.
$w_{i\lambda}$ represents the compatibility relation of the unitary part.
In the cases with $W=0$, we choose one representation of the conjugate pair under the antiunitary operation, and omit the other representation from rows or columns of the differential map matrix. 
The schematic picture describes how representations at ($p-1$)-cells are induced by representations at adjacent $p$-cells.
Black dots denote unitary representations, and $\leftrightarrow$ means the unitary compatibility relation between them. A pair of representations in a red box behaves as a generator of the emergent group.

\subsection{Unitary$\leftrightarrow$unitary}
\underline{$n$=0:}
\begin{align}
[d^1_{p,n}]_{i\lambda}=w_{i\lambda}.
\end{align}
\\
\begin{align}
\begin{array}{c}
\includegraphics[width=4cm,angle=0,clip]{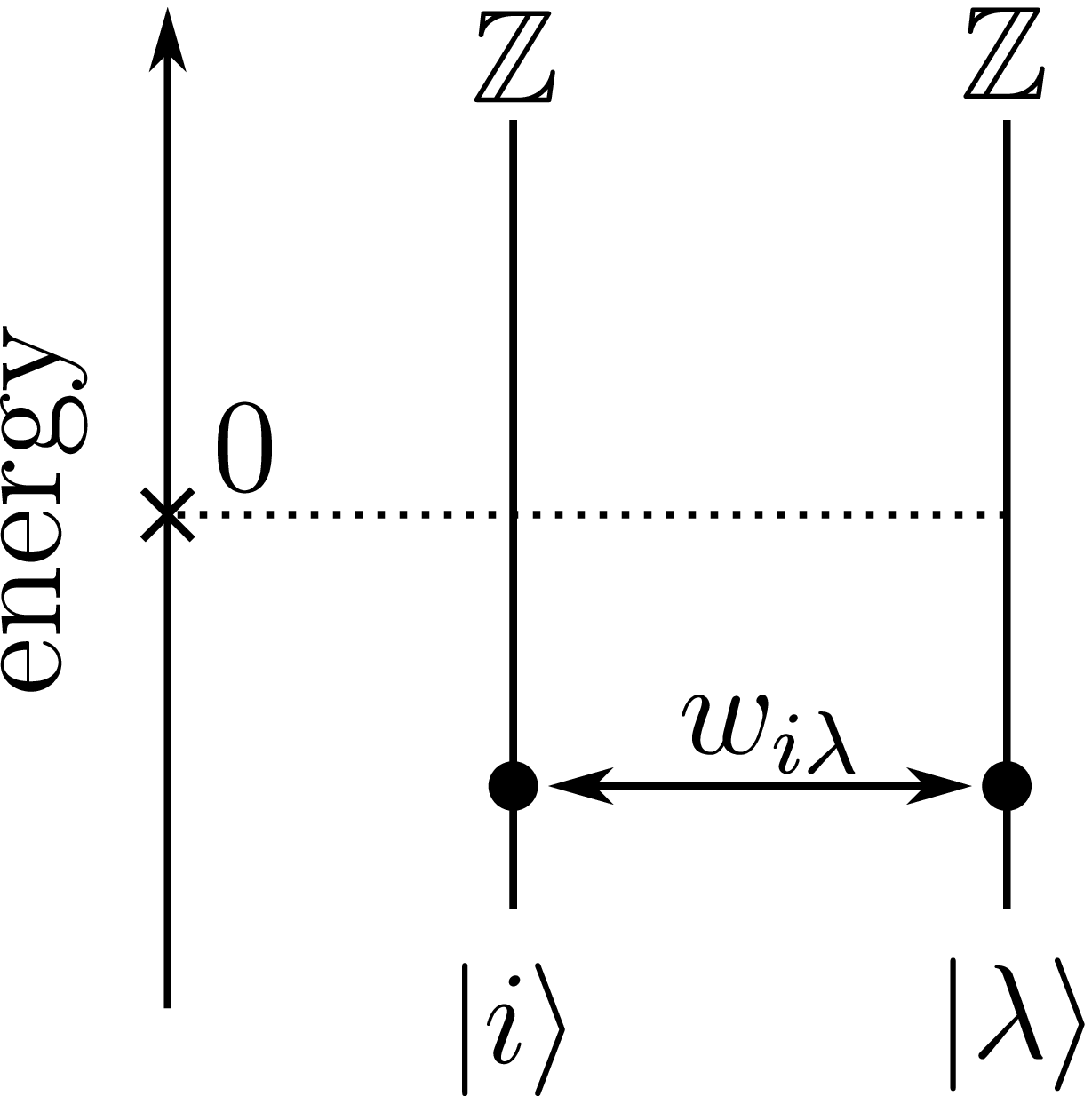}
\end{array}\notag
\end{align}

\subsection{Antiunitary ($W_i=+1$ at $n=0$)$\leftrightarrow$unitary}
\underline{$n$=0:}
\begin{align}
[d^1_{p,n}]_{i\lambda}=2w_{i\lambda}.
\end{align}
\\
\begin{align}
\begin{array}{c}
\includegraphics[width=4cm,angle=0,clip]{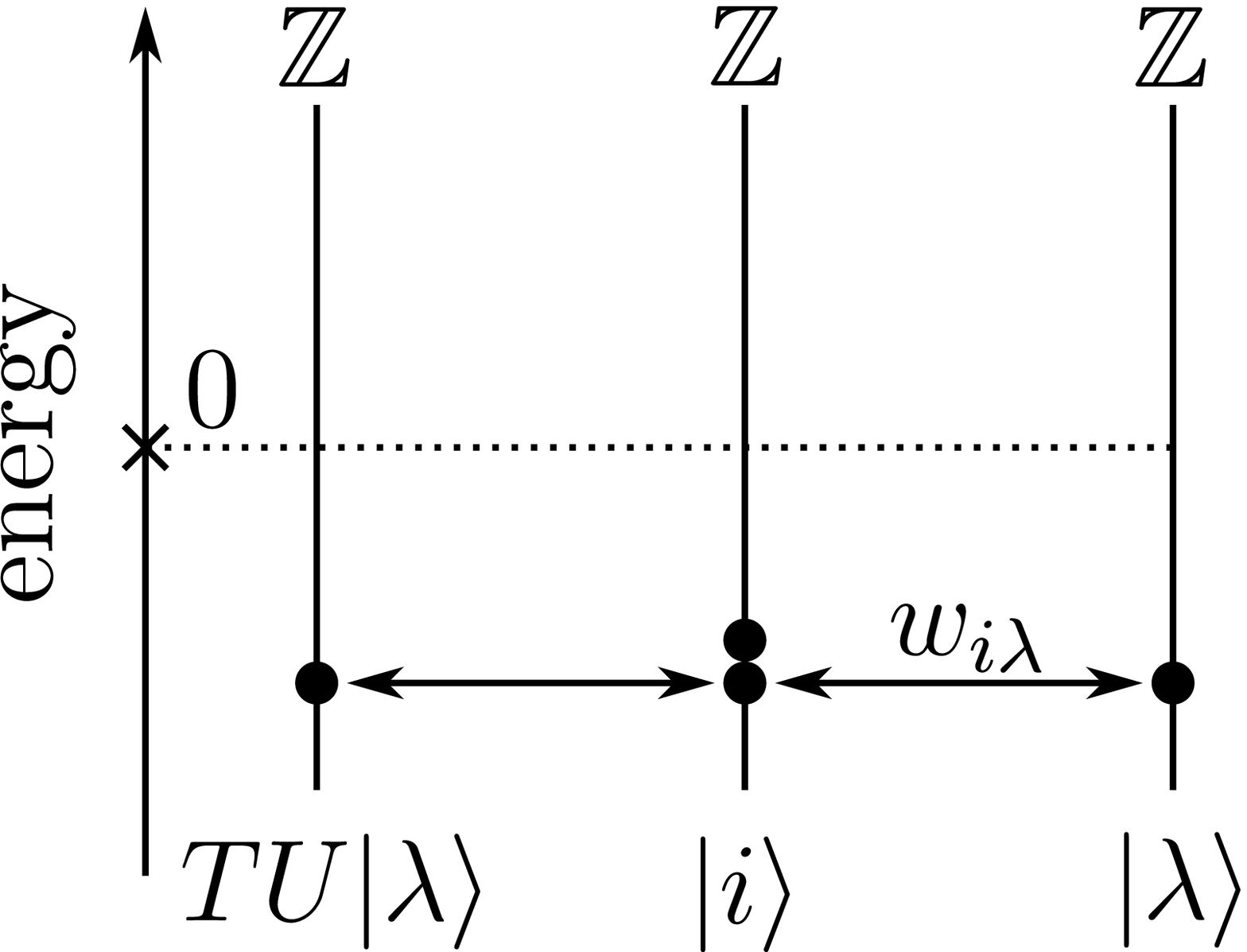}
\end{array}\notag
\end{align}
\underline{$n$=2:}
\begin{align}
[d^1_{p,n}]_{i\lambda}=w_{i\lambda}.
\end{align}
\\
\begin{align}
\begin{array}{c}
\includegraphics[width=4cm,angle=0,clip]{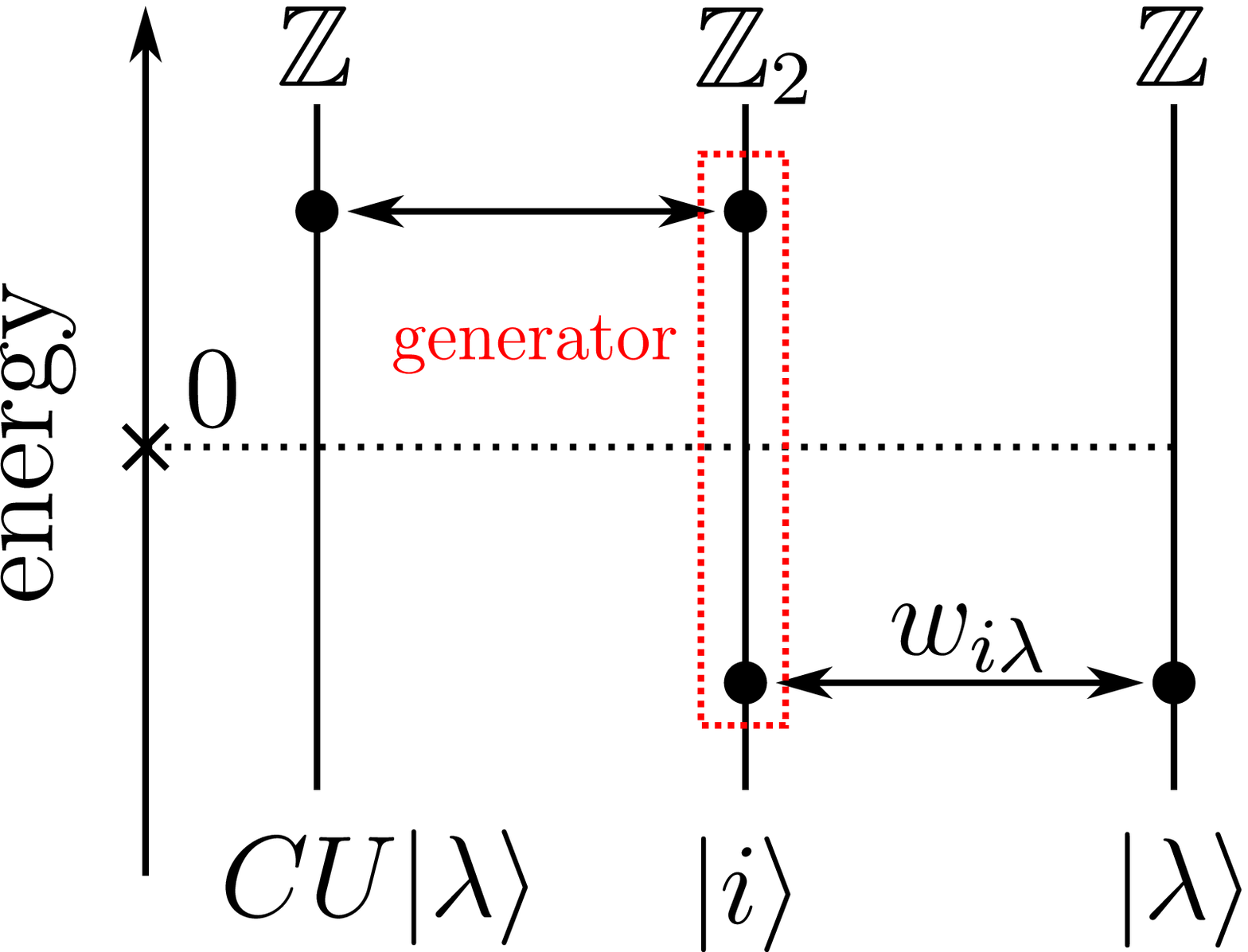}
\end{array}\notag
\end{align}
\underline{$n$=4:}
\begin{align}
[d^1_{p,n}]_{i\lambda}=w_{i\lambda}.
\end{align}
\\
\begin{align}
\begin{array}{c}
\includegraphics[width=4cm,angle=0,clip]{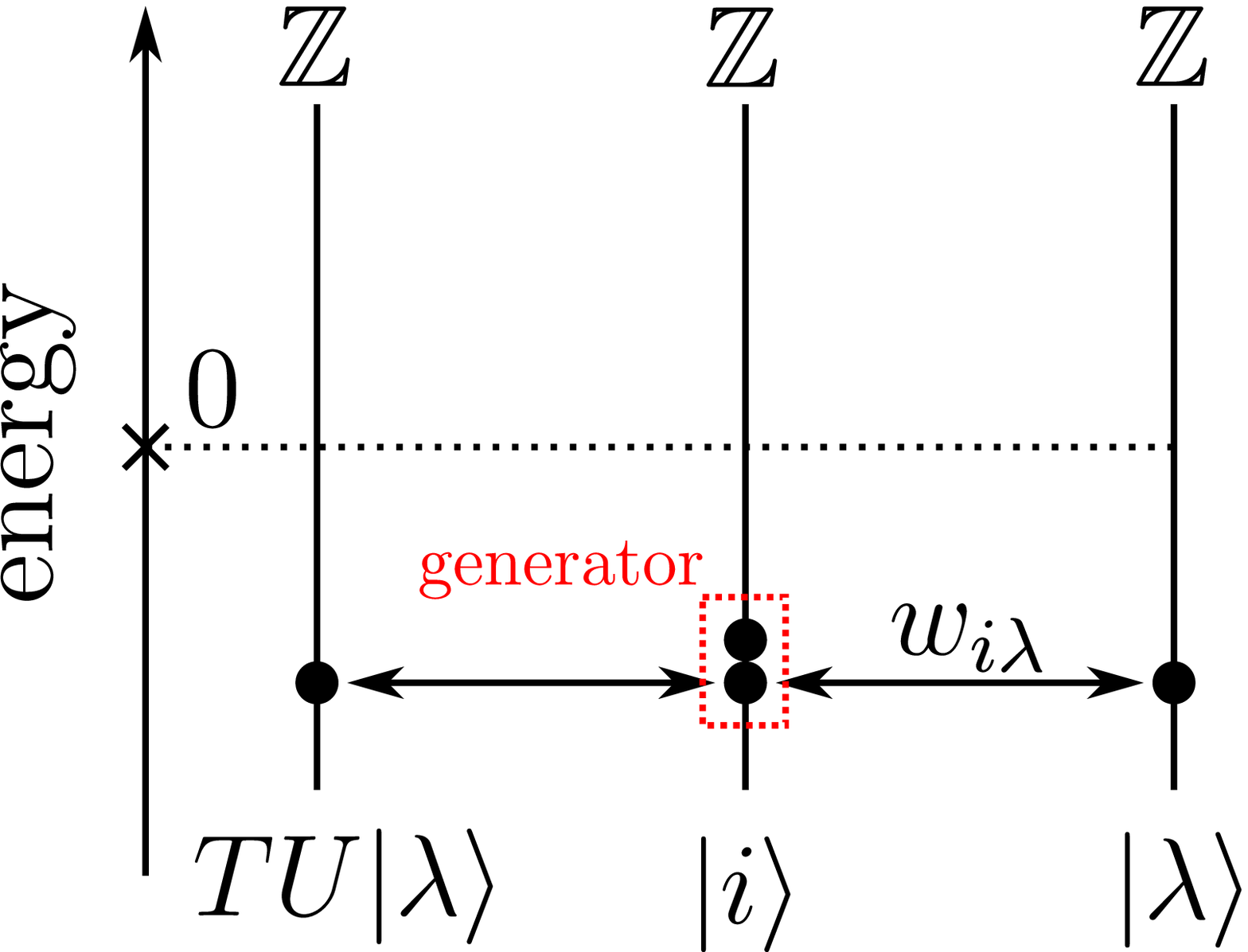}
\end{array}\notag
\end{align}

\subsection{Antiunitary ($W_i=0$ at $n=0$)$\leftrightarrow$unitary}
\underline{$n$=0,4:}
\begin{align}
[d^1_{p,n}]_{i\lambda}=w_{i\lambda}+w_{\overline{i}\lambda}.
\end{align}
\\
\begin{align}
\begin{array}{c}
\includegraphics[width=4cm,angle=0,clip]{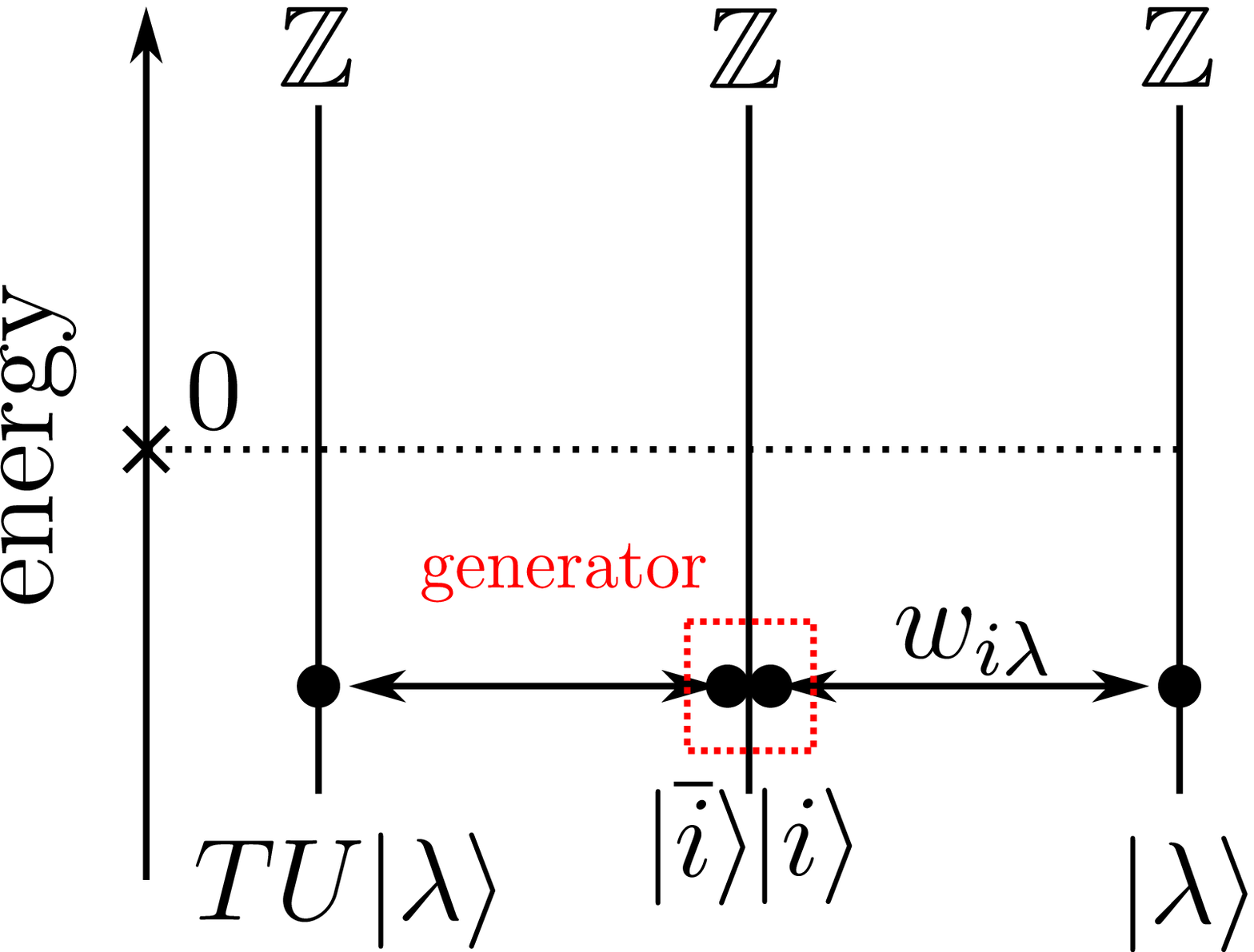}
\end{array}\notag
\end{align}
\underline{$n$=2,6:}
\begin{align}
[d^1_{p,n}]_{i\lambda}=w_{i\lambda}-w_{\overline{i}\lambda}.
\end{align}
\\
\begin{align}
\begin{array}{c}
\includegraphics[width=4cm,angle=0,clip]{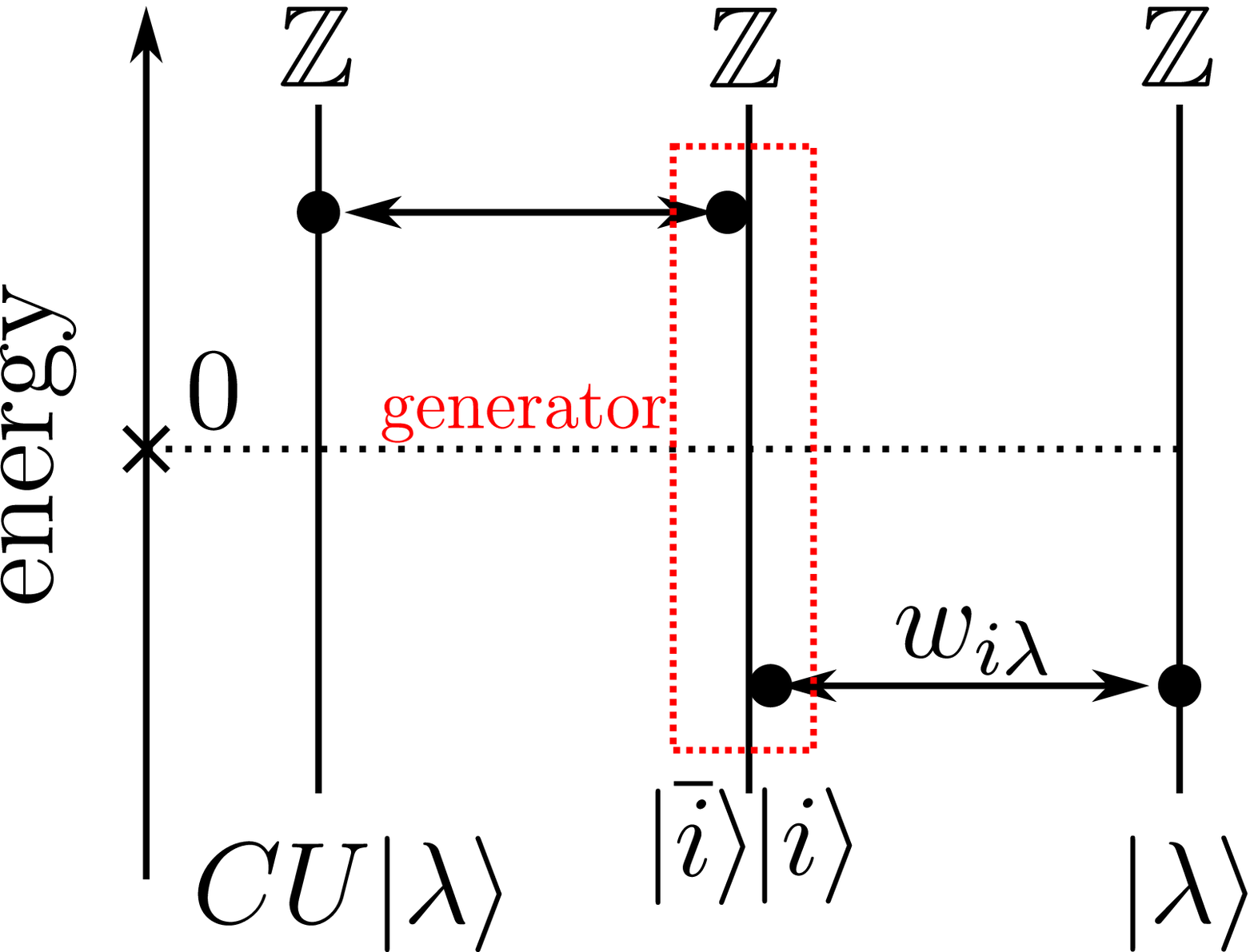}
\end{array}\notag
\end{align}

\subsection{Antiunitary ($W_i=0$ at $n=0$)$\leftrightarrow$antiunitary ($W_\lambda=1$ at $n=0$)}
\underline{$n$=0:}
\begin{align}
[d^1_{p,n}]_{i\lambda}=w_{i\lambda}.
\end{align}
\\
\begin{align}
\begin{array}{c}
\includegraphics[width=4cm,angle=0,clip]{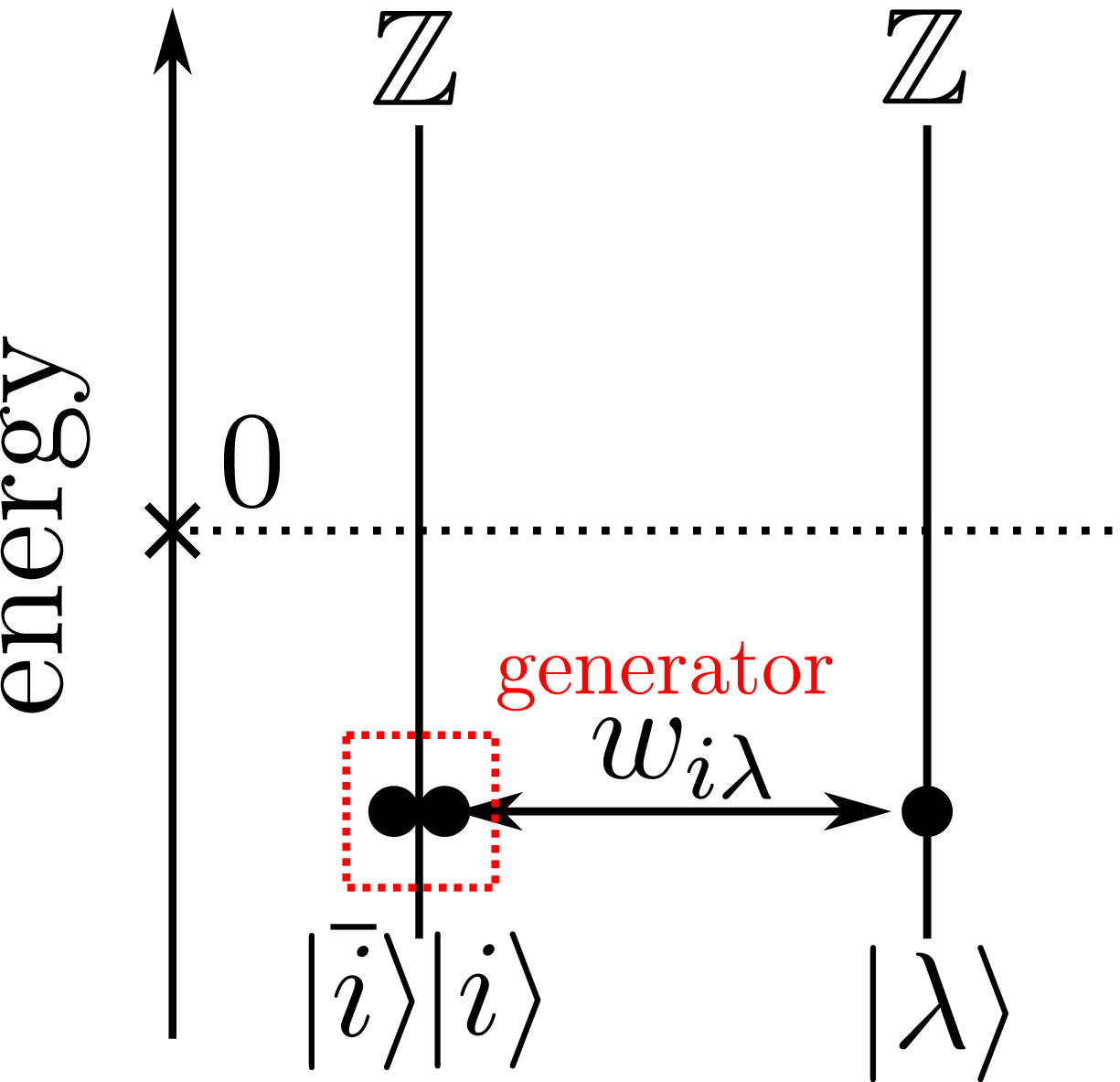}
\end{array}\notag
\end{align}
\underline{$n$=4:}
\begin{align}
[d^1_{p,n}]_{i\lambda}=2w_{i\lambda}.
\end{align}
\\
\begin{align}
\begin{array}{c}
\includegraphics[width=4cm,angle=0,clip]{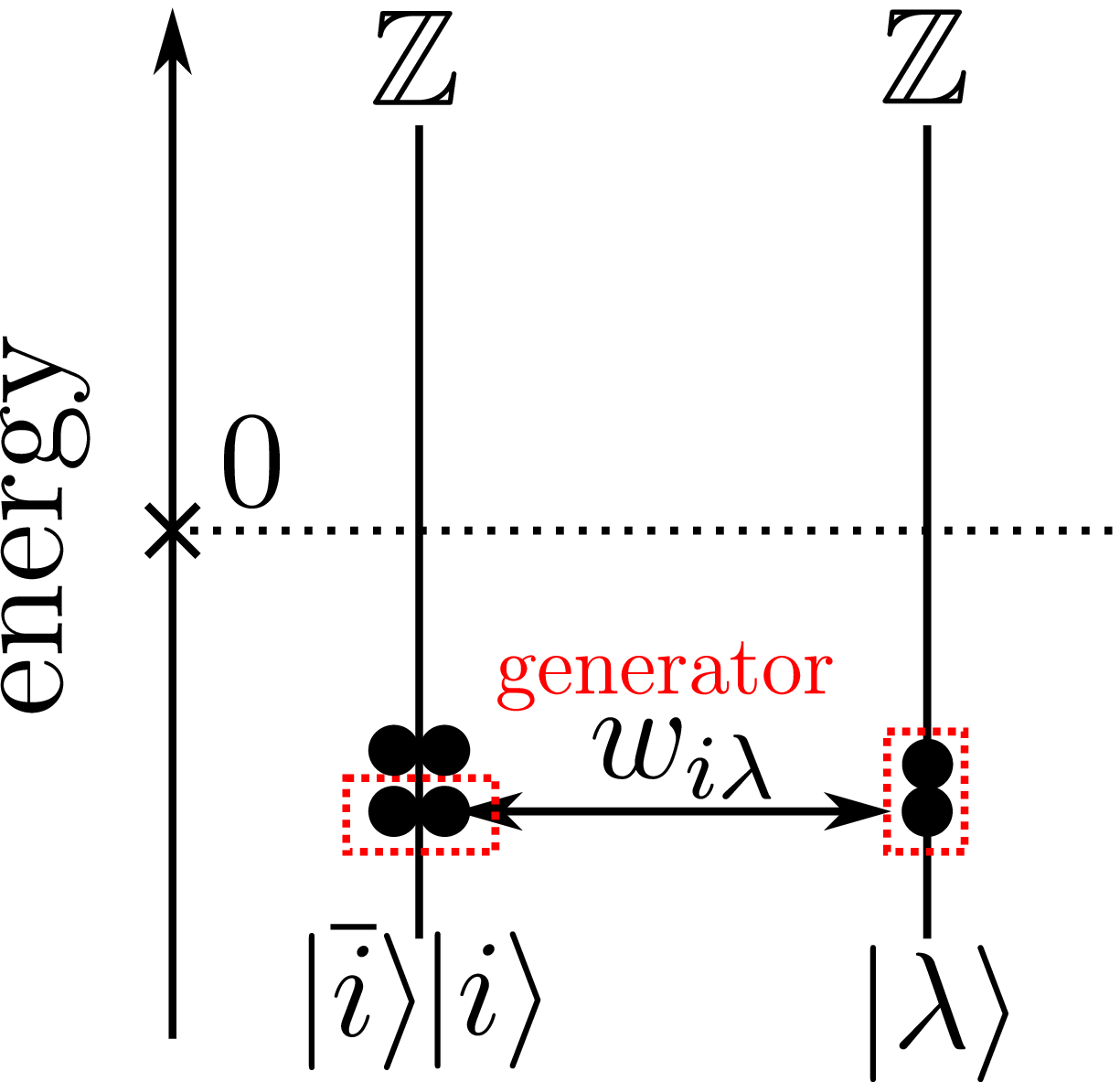}
\end{array}\notag
\end{align}

\subsection{Antiunitary ($W_i=+1$ at $n=0$)$\leftrightarrow$antiunitary ($W_\lambda=0$ at $n=0$)}
\underline{$n$=0:}
\begin{align}
[d^1_{p,n}]_{i\lambda}=2w_{i\lambda}.
\end{align}
\\
\begin{align}
\begin{array}{c}
\includegraphics[width=4cm,angle=0,clip]{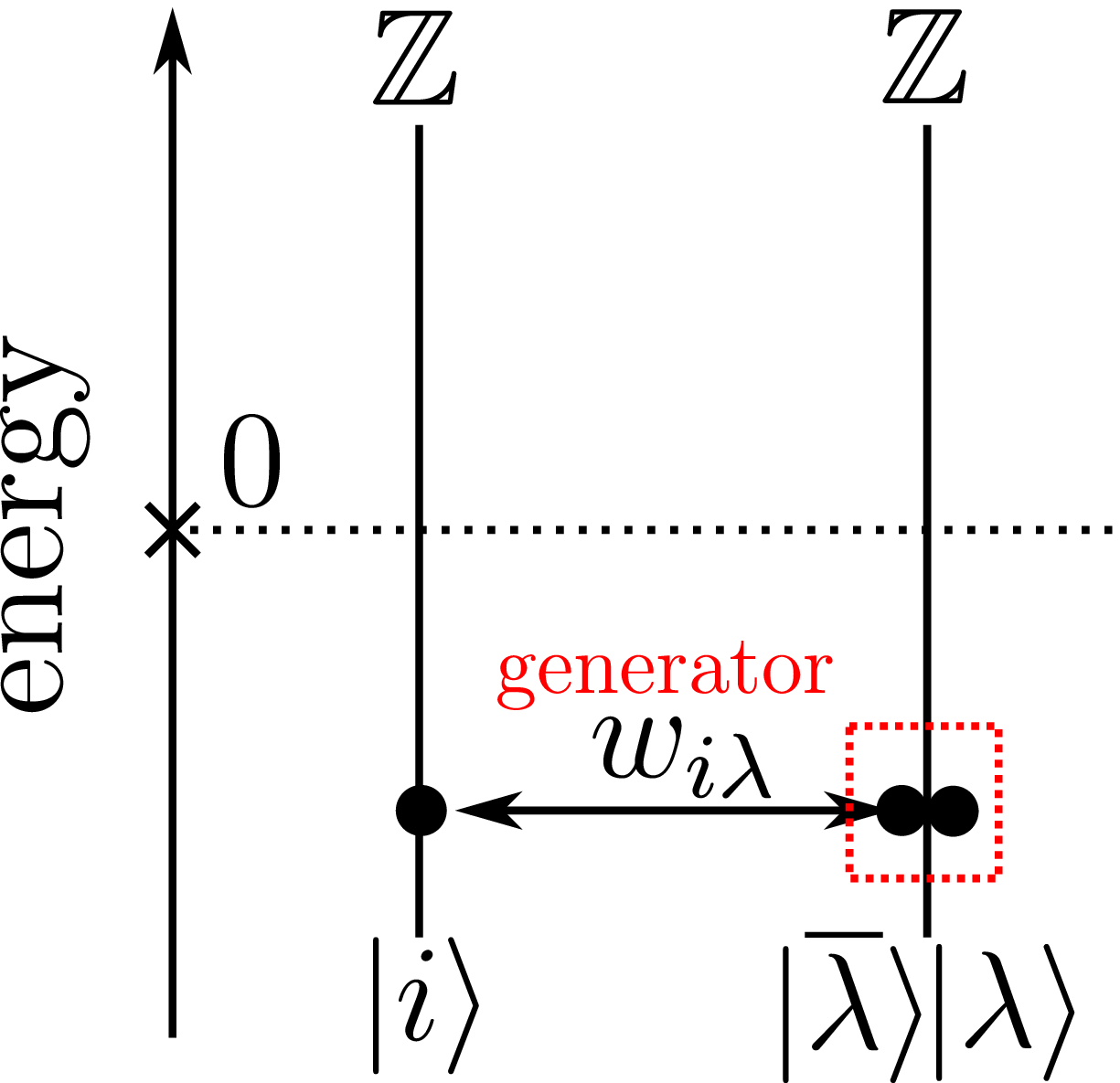}
\end{array}\notag
\end{align}
\underline{$n$=2:}
\begin{align}
[d^1_{p,n}]_{i\lambda}=w_{i\lambda}.
\end{align}
\\
\begin{align}
\begin{array}{c}
\includegraphics[width=4cm,angle=0,clip]{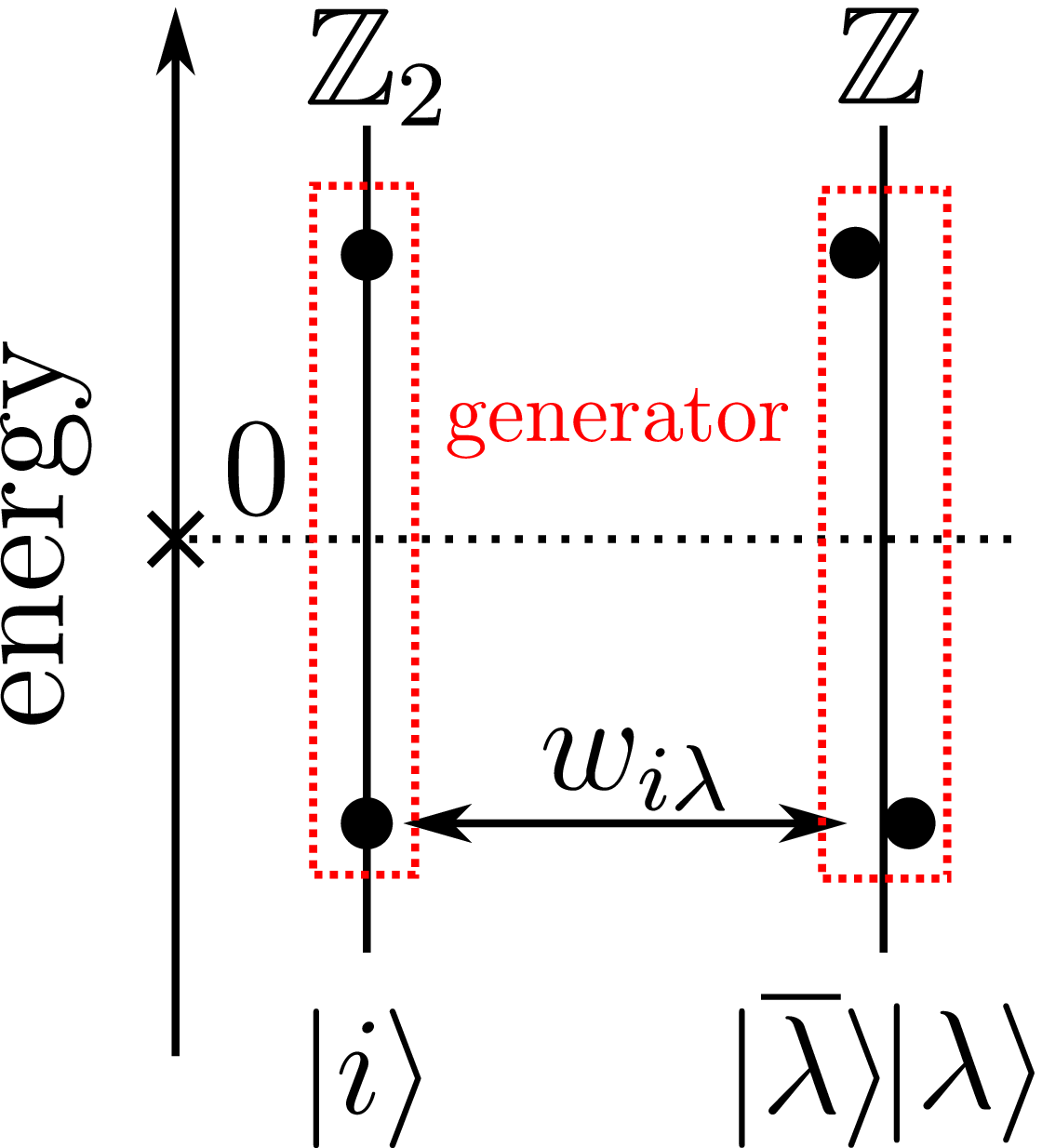}
\end{array}\notag
\end{align}
\underline{$n$=4:}
\begin{align}
[d^1_{p,n}]_{i\lambda}=w_{i\lambda}.
\end{align}
\\
\begin{align}
\begin{array}{c}
\includegraphics[width=4cm,angle=0,clip]{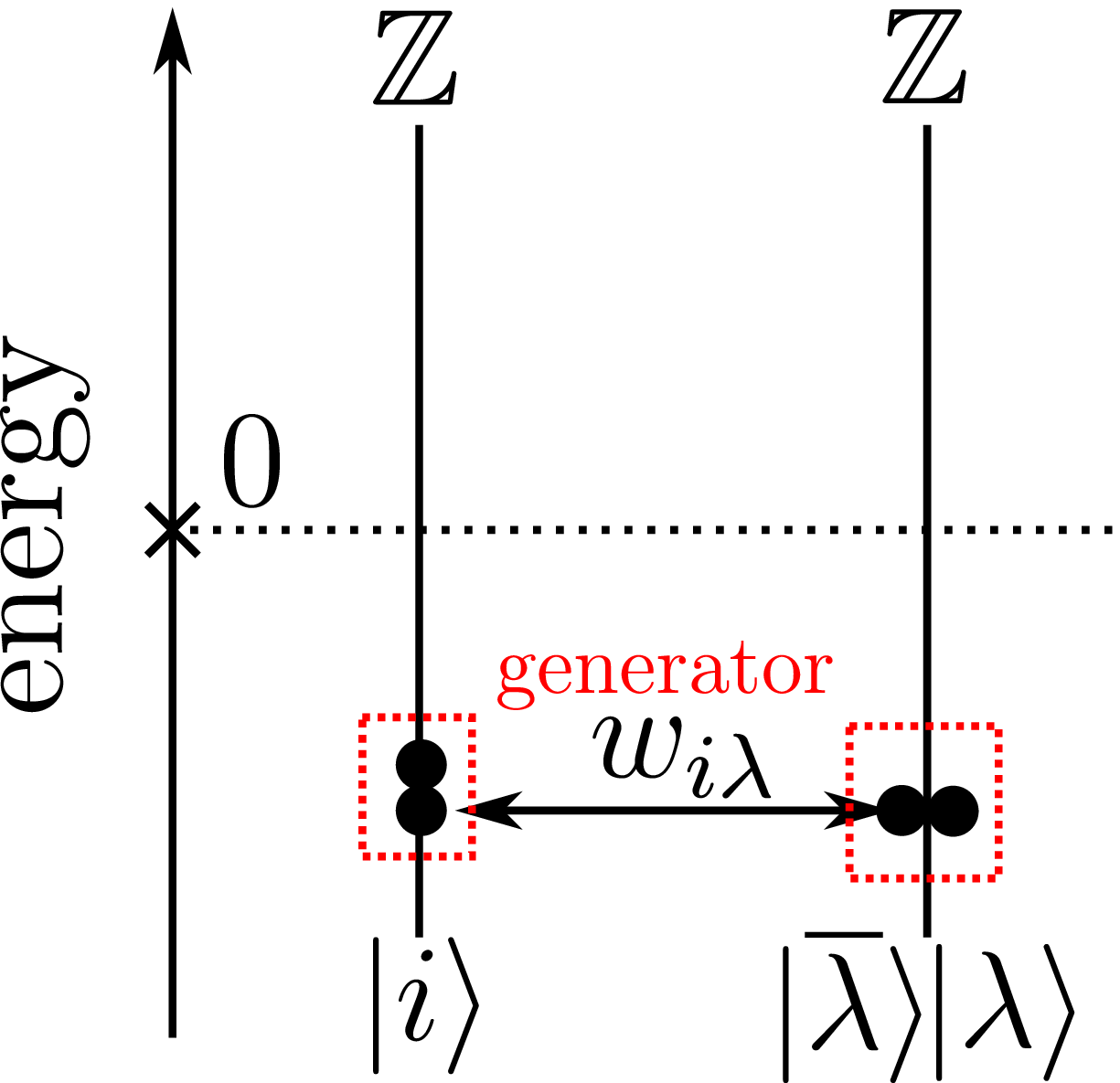}
\end{array}\notag
\end{align}

\subsection{Antiunitary$\leftrightarrow$antiunitary (the same $W$,$W=0$)}

\underline{$n$=0,4:}
\begin{align}
[d^1_{p,n}]_{i\lambda}=w_{i\lambda}+w_{\overline{i}\lambda}.
\end{align}
\\
\begin{align}
\begin{array}{c}
\includegraphics[width=4cm,angle=0,clip]{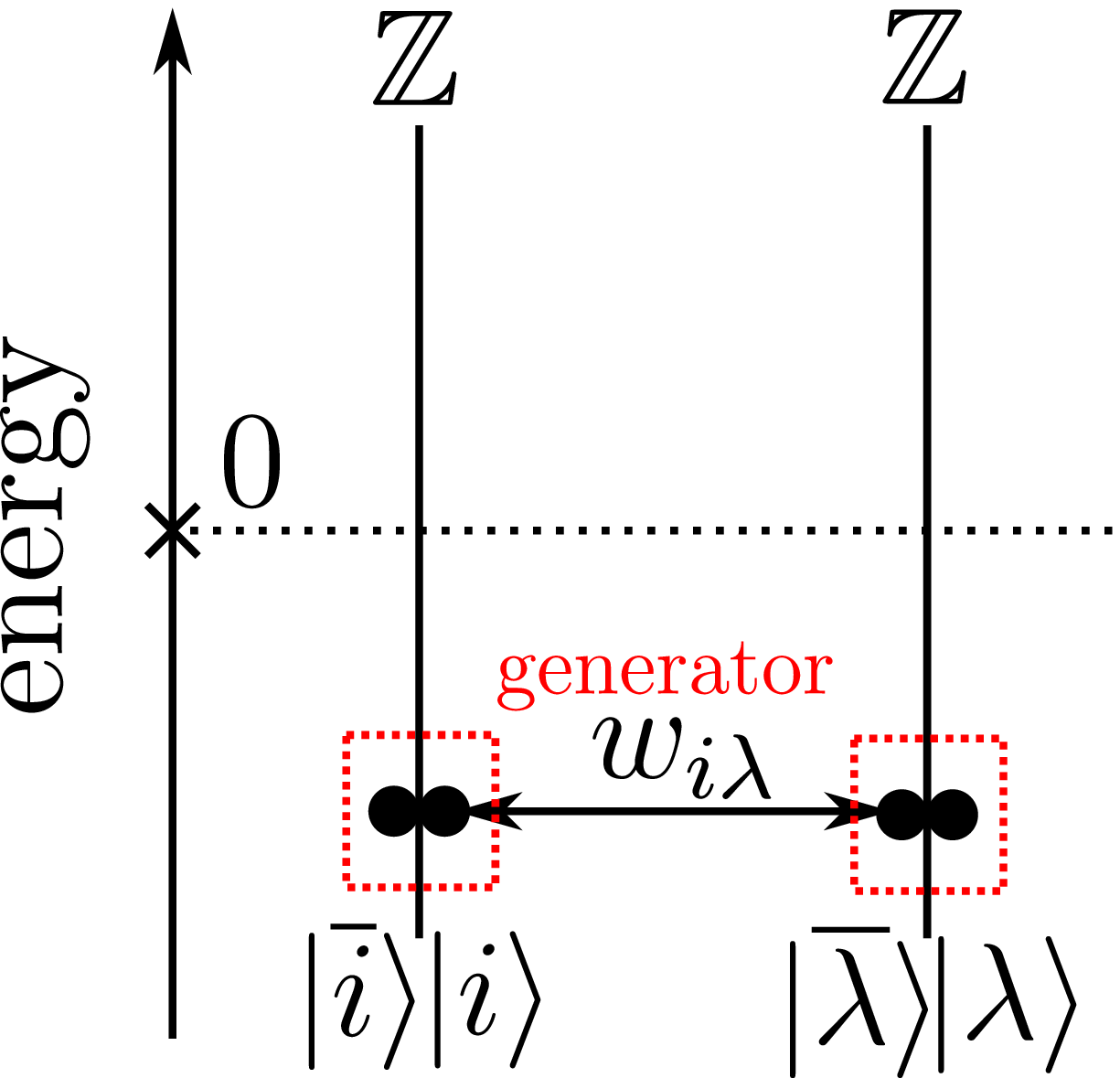}
\end{array}\notag
\end{align}
\underline{$n$=2,6:}
\begin{align}
[d^1_{p,n}]_{i\lambda}=w_{i\lambda}-w_{\overline{i}\lambda}.
\end{align}
\\
\begin{align}
\begin{array}{c}
\includegraphics[width=4cm,angle=0,clip]{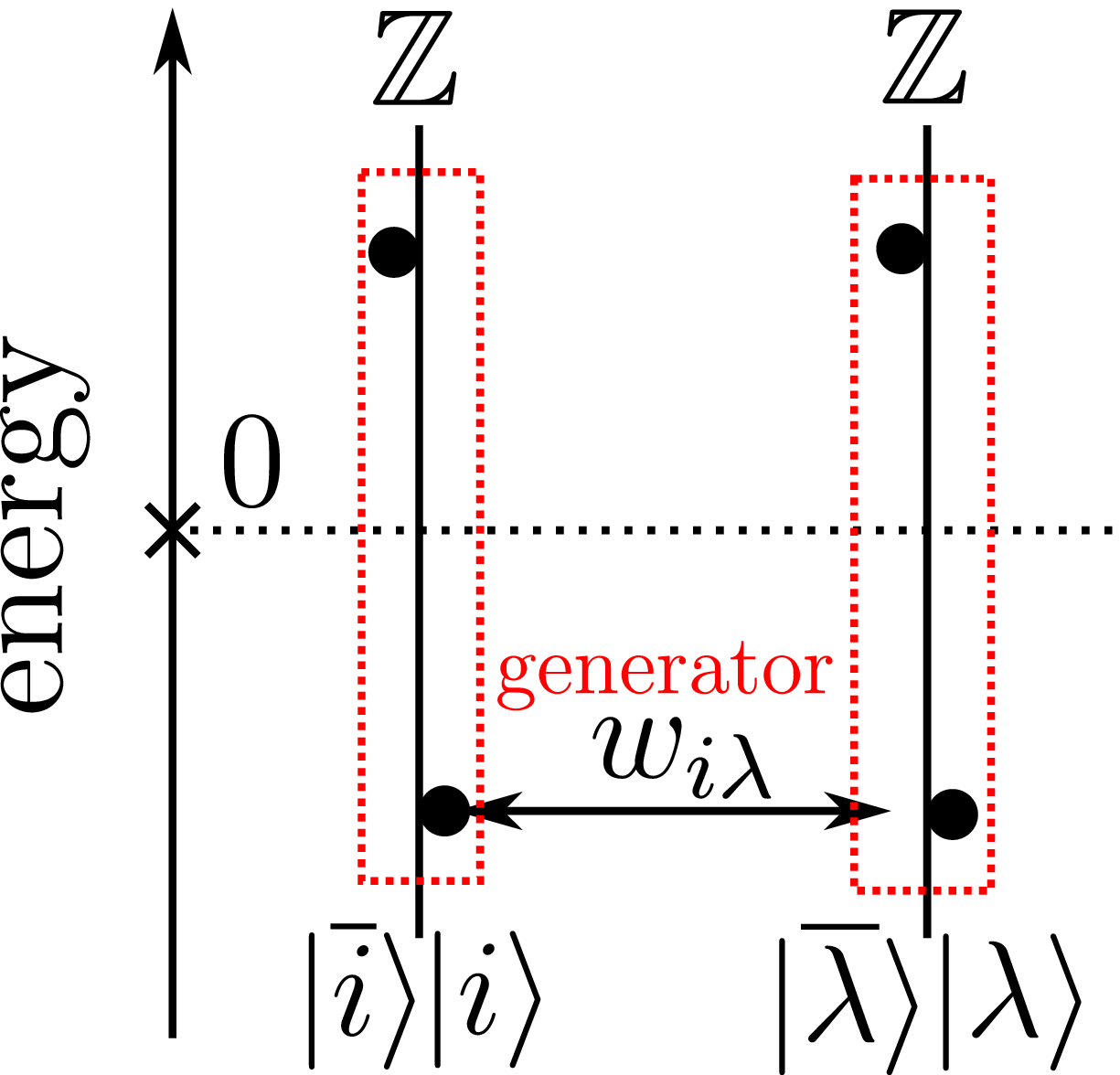}
\end{array}\notag
\end{align}

\subsection{Antiunitary$\leftrightarrow$antiunitary (the same $W$,$W=\pm1$)}
Same as the conventional compatibility relation.

\section{Group extension problems}
Without the time-reversal symmetry (TRS) $\Z_2^T$ in class A systems, because of the absence of a building block state in one and three dimensions, the $K$-group fits into the short exact sequence 
\begin{align}
0 \to E^\infty_{0,0} \to K^G_0(\mathbb{E}^3) \to E^\infty_{2,-2} \to 0. 
\end{align}
When $E^\infty_{2,-2}$ is free Abelian, the above exact sequence splits (as Abelian groups), and the $K$-group is determined as $E^\infty_{0,0} \oplus E^\infty_{2,-2}$. 
For nonmagnetic or magnetic point group symmetries in class A systems, from the explicit computation of the AHSS, we find that group extensions in which $E^\infty_{2,-2}$ contains a torsion Abelian group take the following form 
\begin{align}
0 \to E^\infty_{0,0} \to K^G_0(\mathbb{E}^3) \to \underbrace{\Z_2}_{E^\infty_{2,-2}} \to 0, 
\label{eq:app_group_extension}
\end{align}
where $E^\infty_{2,-2} = \Z_2$ represents second-order topological insulators (TIs) and generated by embedding Chern insulators on 2-cells with respecting the (non)magnetic point group symmetry. 
The purpose of this Appendix is to depict how we solve the group extension (\ref{eq:app_group_extension}) systematically. 

\subsection{Equivalence of $\Z_2$ second-order TIs and the three-dimensional TI with mass domain walls}
In the following manner, one can avoid solving $ad$ $hoc$ problems that depend on what the (non)magnetic point group is. 
Notice that the second-order TIs generating $E^\infty_{2,-2} = \Z_2$ can be represented by the $4\times 4$ three-dimensional Dirac Hamiltonian with the $O(3) \times \Z_2^T$ symmetry 
\begin{align}
&H(\bm{x}) = -i \p_x \sigma_x \tau_x - i \p_y \sigma_y \tau_x -i \p_z \sigma_z \tau_x + m \tau_z, \label{eq:app_4by4_dirac} \\
&IH(\bm{x})I^{-1} = H(-\bm{x}), I=\tau_z, \label{eq:app_4by4_dirac_i} \\
&C_{\bm{n},\theta} H(\bm{x}) (C_{\bm{n},\theta})^{-1} = H(\hat R_{\bm{n},\theta}\bm{x}), C_{\bm{n},\theta} = e^{-i \frac{\theta}{2} \bm{n} \cdot \bm{\sigma}}, \label{eq:app_4by4_dirac_r} \\
&TH(\bm{x})T^{-1}=H(\bm{x}),  T=i\sigma_y K, \label{eq:app_4by4_dirac_t} 
\end{align}
where $\bm{x} \mapsto R_{\bm{n},\theta} \bm{x}$ is the rotation around the $\bm{n}$-axis by the angle $\theta$, and $K$ is the complex-conjugate operator. 
In the presence of the pure TRS $T$, the mass term $M(\bm{x}) \tau_y$ is forbidden at any spatial point, which is nothing but the robustness of the time-reversal symmetric three-dimensional TI. 
Let us consider a (non)magnetic point group symmetry $G \subset O(3) \times \Z_2^T$ where $G$ does not include the pure TRS. 
Owing to the absence of the TRS, without breaking the symmetry $G$, it is possible to add a spatially varying mass term $M(\bm{x}) \tau_y$ to the Hamiltonian (\ref{eq:app_4by4_dirac}). 
If the symmetry $G$ forbids the mass $M(\bm{x})$ to be uniform, a possible spatial configuration of $M(\bm{x})$ should accompany with symmetry-respecting domain walls at which the sign of $M(\bm{x})$ changes, resulting in the Chern insulators with a unit Chern number localized at the domain walls. 
We find that any second-order TIs with $E^\infty_{2,-2}=\Z_2$ can be represented in this way. 

We present an example.
For the nonmagnetic point group $\bar 1 = \{1,I\}$, the inversion symmetry enforces a domain wall of $M(\bm{x})$ at the origin. 
In the presence of a single domain wall along the $z$-direction with $M(-z) = - M(z)$ so that $M(z \to \infty)>0$, the low-energy states are described by the doublet $\Phi(z)$ localized at the domain wall ($z \sim 0$) 
\begin{align}
\Phi_{\rm loc}(z) 
= \left( 
\begin{pmatrix}1\\0\\\end{pmatrix}_\sigma \otimes
\begin{pmatrix}0\\1\\\end{pmatrix}_\tau , 
\begin{pmatrix}0\\1\\\end{pmatrix}_\sigma \otimes
\begin{pmatrix}1\\0\\\end{pmatrix}_\tau 
\right) \notag\\
\times e^{-\int^zM(z') dz'}. 
\end{align}
The low-energy effective two-dimensional Hamiltonian $H_{2d}$ is given as 
\begin{align}
&H \Phi_{\rm loc}(z) = \Phi_{\rm loc}(z) H_{2d}, \\
&H_{2d}
= 
-i \p_x \sigma_x - i \p_y \sigma_y + m \sigma_z.
\end{align}
The two-dimensional Hamiltonian $H_{2d}$ represents the Chern insulator with a unit Chern number, a representative of the second-order TI phase with $E^\infty_{2,-2}=\Z_2$.

\subsection{Uniform mass and triviality of group extension}
Now we solve the group extension (\ref{eq:app_group_extension}).
We discuss whether or not the stack of two $\Z_2$ second-order TIs is adiabatically equivalent to a generator of $E^\infty_{0,0}$, a zero-dimensional bound state at the origin. 
Along the line of the thought in the previous subsection, instead of starting with Chern insulators on 2-cells, we consider a stack of two three-dimensional TIs 
\begin{align}
H(\bm{x})^{\oplus 2}
=
(-i \p_x \sigma_x \tau_x - i \p_y \sigma_y \tau_x -i \p_z \sigma_z \tau_x + m \tau_z) \otimes \mu_0, 
\end{align}
where $\mu$s are Pauli matrices for layers. 
We have four mass terms $M \tau_y, M_1 \mu_x \tau_y, M_2 \mu_y \tau_y$ and $M_3 \mu_z \tau_y$. 
Note that $M \tau_y$ commutes with the others. 
Among them, we can exclude $M \tau_y$ since we have assumed nontrivial $\Z_2$ second-order TI phases where there is no uniform mass term of $M \tau_y$, and $M \tau_y$ cannot contribute to the construction of a zero-dimensional bound state. 
Let us introduce the mass vector $\bm{M}(\bm{x}) = (M_1(\bm{x}),M_2(\bm{x}),M_3(\bm{x}))$. 
From the group actions (\ref{eq:app_4by4_dirac_i}), (\ref{eq:app_4by4_dirac_r}) and (\ref{eq:app_4by4_dirac_t}), the mass vector $\bm{M}(\bm{x})$ changes as in Eq. ($\ref{eq:condition_m_vector}$) under the (non)magnetic point group. 
We find a simple criterion: 
\begin{thm}
The group extension (\ref{eq:app_group_extension}) is trivial if and only if there exists a uniform mass vector $\bm{M}(\bm{x}) = \bm{M}_0$. 
\label{app:thm}
\end{thm}
The proof is as follows. 
The ``if'' part is obvious: a uniform mass vector induces a large mass gap in the whole real space, leading to the absence of any low-energy state with the same energy scale as $m$. 
The ``only if'' part is more involved. 
Let $G$ be a (non)magnetic point group. 
Suppose a $G$-symmetric hedgehog of the mass vector $\bm{M}(\bm{x})$ with a finite winding number $q = \frac{1}{4 \pi} \int_{|\bm{x}| \to \infty} \hat M \cdot (d \hat M \times d \hat M) \in \Z$ with $\hat M(\bm{x}) = \bm{M}(\bm{x})/|\bm{M}(\bm{x})|$. 
Possible winding numbers $q \in \Z$ are restricted to some set of integers from the symmetry $G$. 
Applying the index theorem for an infinite open manifold\cite{callias,weinberg,niemi} to the defect Hamiltonian $H(\bm{x})^{\oplus 2} + \bm{M}(\bm{x}) \cdot \bm{\mu}\tau_y$, we find that there appear $q$ stable zero-dimensional bound states $\{\psi_j\}_{j=1}^q$, and the low-energy effective Hamiltonian $H_{0d}$ describing them is $H_{0d} = m \bm{1}_{q \times q}$. 
The triviality of the group extension (\ref{eq:app_group_extension}) implies that the bound states $\{\psi_j\}_{j=1}^q$ belong to the image of the first differential $d^1_{1,0}$. 
This is equivalent to that all the bound states $\{\psi_j\}_{j=1}^q$ can spatially split at the origin and move far away (See Fig. \ref{fig3}(a)). 
In doing so, the mass vector around the origin becomes finite and can be uniform (because of the absence of the winding number near the origin). 
This completes the proof\cite{theoremb}. 


Let us see a few examples.
\begin{itemize}
\item[$\bar 1$:]
The inversion symmetry imposes the condition $\bm{M}(-\bm{x})=-\bm{M}(\bm{x})$ on the mass vector. 
There is no uniform mass vector, meaning that the group extension is nontrivial. 
\item[$\bar 4$:]
The rotoinversion symmetry $IC^z_4$ imposes the condition $\bm{M}(y,-x,-z)=-\bm{M}(x,y,z)$ on the mass vector. 
There is no uniform mass vector, meaning that the group extension is nontrivial. 
\item[$4'$:]
The time-reversal rotation symmetry $TC^z_4$ induces the constraint $(-M_1,M_2,-M_3)|_{-y,x,z}=\bm{M}(x,y,z)$.
There exists a uniform mass vector $\bm{M}(x,y,z)=(0,M_0,0)$, meaning that the group extension is trivial. 
\end{itemize}

Using Theorem \ref{app:thm}, it is easy to determine if the group extension (\ref{eq:app_group_extension}) is trivial or not for any nonmagnetic and magnetic point group symmetries. 
The results are summarized in Figs. \ref{fig4} and \ref{fig6}.

\subsection{Some analytic solutions}
For some simple (non)magnetic point group symmetries, one can derive the explicit zero mode solution for a given winding number $q$. 
When the mass vector $\bm{M}(x,y,z)$ is composed of a single domain wall of the third component $M_3(z)$ along the $z$-direction and a vortex line of the first and second components $(M_1(r,\theta),M_2(r,\theta))$ in the $xy$-plane as in 
\begin{align}
&\bm{M}(r,\theta,z)
= 
(\Delta(r) \cos(q \theta), \Delta(r) \sin (q \theta), m(z)),  q \in \Z, \\
&\Delta(r)>0, \Delta(r \to 0) = 0, \Delta(r\to \infty) = \Delta_0, \\
&m(z \to \pm \infty) = \pm m_0, 
\end{align}
it is straightforward to get the analytic solution of zero modes of the defect Hamiltonian\cite{fukui}  
\begin{align}
\wt H(\bm{x})
= 
-i \bm{\p} \cdot \bm{\sigma} \tau_x + \bm{M}(\bm{x}) \cdot \bm{\mu} \tau_y 
\end{align}
with the chiral symmetry $\tau_z$, $\{\wt H(\bm{x}),\tau_z\}=0$. 
Without loss of generality, one can assume $m_0>0$. 
From the $U(1)$ rotation symmetry generated by the angular momentum $J_z$ (see below) which commutes with the chiral symmetry $\tau_z$, the zero modes are simultaneously labeled by the chirality $\tau_z = \pm 1$ and the angular momentum $j_z$. 
The explicit forms of the zero modes are given as 
\begin{align}
&\tau_z \psi_{j_z}^\pm(r,\theta,z)=\pm \psi_\pm(r,\theta,z), \\
&J_z \psi^\pm_{j_z}(r,\theta,z)={j_z}\psi^\pm_m(r,\theta,z),  J_z=-i \p_\theta+\frac{1}{2}(\sigma_z+q\mu_z), 
\end{align}
\begin{align}
{j_z} \in \left\{\begin{array}{ll}
\Z+\frac{1}{2} & (q\in 2\Z), \\
\Z & (q\in 2\Z+1), \\
\end{array}\right. , 
|{j_z}|<
\left\{\begin{array}{ll}
\frac{1+q}{2} & (\tau_z=1), \\
\frac{1-q}{2} & (\tau_z=-1),\\
\end{array}\right.
\end{align}
\begin{widetext}
\begin{align}
\psi^+_{j_z}(r,\theta,z) 
\sim e^{i({j_z}-\frac{\sigma_z+q\mu_z}{2}) \theta} 
\left( 
\alpha^+_{j_z}(r) 
\begin{pmatrix}1\\0\\\end{pmatrix}_{\sigma}\otimes
\begin{pmatrix}0\\1\\\end{pmatrix}_{\mu}\otimes
\begin{pmatrix}1\\0\\\end{pmatrix}_{\tau}
+
\beta^+_{j_z}(r) 
\begin{pmatrix}0\\1\\\end{pmatrix}_{\sigma}\otimes
\begin{pmatrix}1\\0\\\end{pmatrix}_{\mu}\otimes
\begin{pmatrix}1\\0\\\end{pmatrix}_{\tau}
\right)
e^{-\int^z m(z')dz'}, 
\label{eq:app_zero_+}
\end{align}
\begin{align}
\psi^-_{j_z}(r,\theta,z) 
\sim e^{i({j_z}-\frac{\sigma_z+q\mu_z}{2}) \theta} 
\left( 
\alpha^-_{j_z}(r) 
\begin{pmatrix}1\\0\\\end{pmatrix}_{\sigma}\otimes
\begin{pmatrix}1\\0\\\end{pmatrix}_{\mu}\otimes
\begin{pmatrix}0\\1\\\end{pmatrix}_{\tau}
+
\beta^-_{j_z}(r) 
\begin{pmatrix}0\\1\\\end{pmatrix}_{\sigma}\otimes
\begin{pmatrix}0\\1\\\end{pmatrix}_{\mu}\otimes
\begin{pmatrix}0\\1\\\end{pmatrix}_{\tau}
\right)
e^{-\int^z m(z')dz'}.
\label{eq:app_zero_-}
\end{align}
\end{widetext}
Here, the functions $\alpha^\pm_{j_z}(r)$ and $\beta^\pm_{j_z}(r)$ are determined by the detail of the amplitude $\Delta(r)$. 
The relationship among the winding number $q$ and the quantum numbers $\{\tau_z,j_z\}$ is summarized as 
\begin{align}
&q>0, q \in 2\Z  \Rightarrow  \tau_z=1, {j_z}=\pm \frac{1}{2}, \pm \frac{3}{2},\dots, \pm \frac{q-1}{2}, \\
&q>0, q \in 2\Z+1  \Rightarrow  \tau_z=1, {j_z}=0,\pm 1, \pm 2,\dots, \pm \frac{q-1}{2}, \\
&q<0, q \in 2\Z  \Rightarrow  \tau_z=-1, {j_z}=\pm \frac{1}{2}, \pm \frac{3}{2},\dots, \pm \frac{|q|-1}{2}, \\
&q<0, q \in 2\Z+1  \Rightarrow  \tau_z=-1, {j_z}=0,\pm 1, \pm 2,\dots, \pm \frac{|q|-1}{2}.
\end{align}

Using the explicit forms (\ref{eq:app_zero_+}) and (\ref{eq:app_zero_-}) of zero modes, for some cases, one can explicitly obtain representations of zero modes under the (non)magnetic point group symmetry, i.e.\ the element of $E^1_{0,0}$. 
Let us see a few examples.
\begin{itemize}
\item[$\bar 1$:]
In this case, $q$ is constrained into odd integers $q \in 2\Z+1$ from the inversion symmetry $I$. 
For $q>0$, there are $q$ zero modes $\{\psi^+_{j_z}\}_{j_z=-(q-1)/2}^{(q-1)/2}$ with the positive chirality, and these have the inversion eigenvalues $I \psi^+_{j_z}(r,\theta+\pi,-z) = (-1)^{{j_z}+(q-1)/2} \psi^+_{j_z}(r,\theta,z)$. 
A pair of inversion eigenvalues $I=\{1,-1\}$ is in the image of $d^1_{1,0}$. 
Therefore, for any odd integer $q$, the set of bound states $\{\psi^+_{j_z}\}_{j_z=-(q-1)/2}^{(q-1)/2}$ belongs to the generator of $E^2_{0,0}=\Z$. 

\item[$\bar 4$:]
In this case, $q$ is in $4\Z+2$ from the rotoinversion symmetry $IC^z_4$. 
For $q>0$, there exist $q$ zero modes with the positive chirality, and these have the $IC^z_4$ eigenvalues as $I C^z_4\psi^+_{j_z}(r,\theta+\pi/2,-z) = e^{i \pi {j_z}/2} (-1)^{(q-2)/4} \psi^+_{j_z}(r,\theta,z)$. 
Per the increment $q \mapsto q+4$, there appears a quartet of irreducible representations with $IC^z_4 = \{e^{\pi i/4}, e^{3\pi i/4}, e^{5 \pi i/4}, e^{7 \pi i/4}\}$, which is trivial in the sense of being in the image of $d^1_{1,0}$. 
Therefore, for any integer $q \in 4 \Z+2$, there remain two irreducible representations with $IC^z_4 = \{e^{\pi i/4}, e^{3\pi i/4}\}$, the generator of $E^2_{0,0}=\Z^2$. 
\end{itemize}

\section{Schematic pictures of second-order topological insulators}
\label{sec:app_c}
\begin{figure*}[]
\begin{center}
　　　\includegraphics[width=16cm,angle=0,clip]{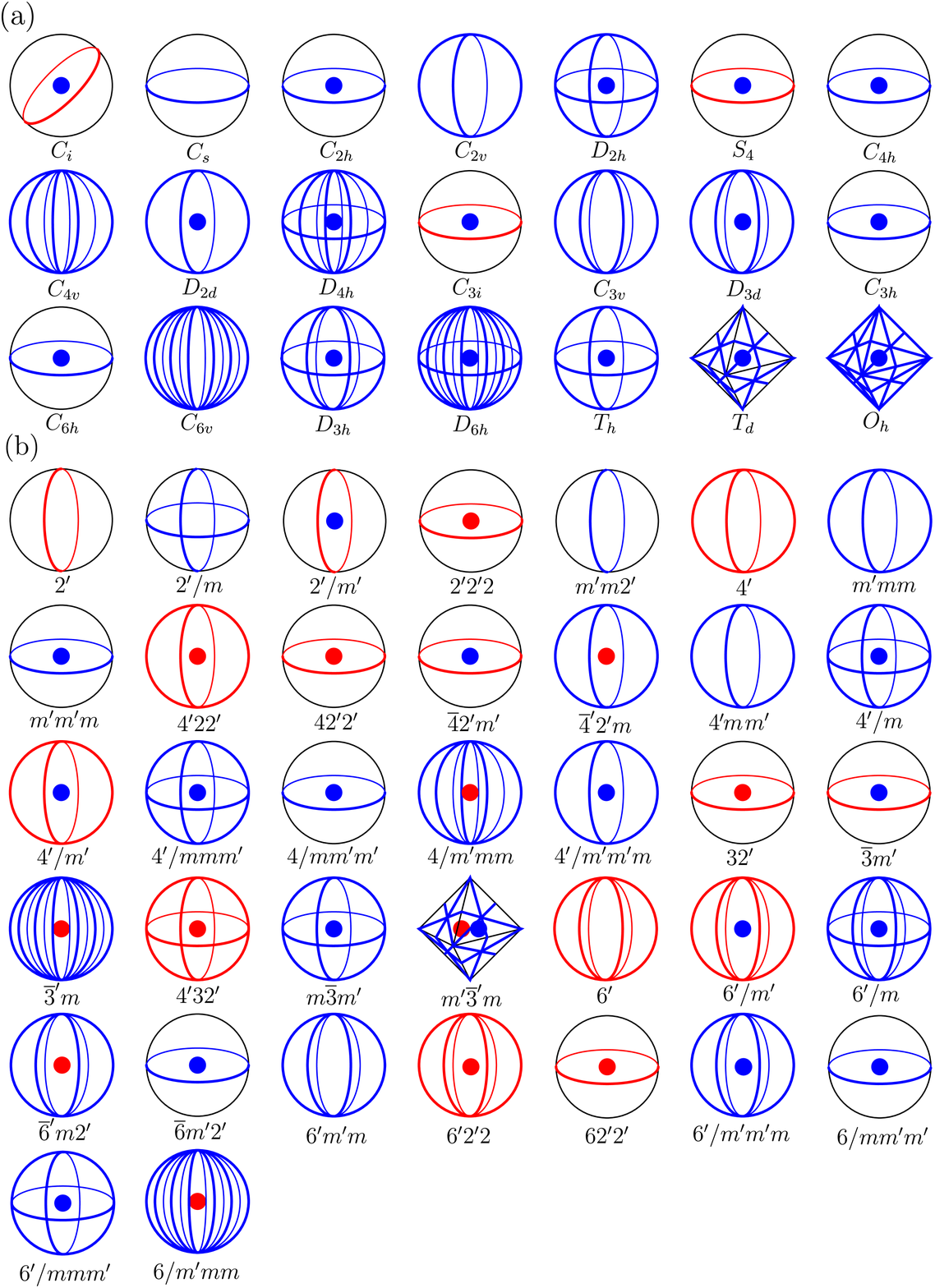}
　　　\caption{Nontrivial higher-order topological states of spinful fermions under (a) nonmagnetic and (b) magnetic point groups. Blue and red objects represent the $\Z^m$ and $\Z_2^n$ ($m,n$: given in Figs.\ref{fig4} and \ref{fig6}) , respectively. The surface edge modes represent second-order topological insulators $E^\infty_{2,-2}$, while the bound states at the origin represent the fourth-order topological insulators $E^\infty_{0,0}$. }
　　　\label{fig8}
\end{center}
\end{figure*}

Locations of surface edge states in second-order topological insulators and bound states at the origin are shown in Fig. $\ref{fig8}$.
The corresponding classifications are given in Figs. $\ref{fig4}$ and $\ref{fig6}$.

\section{$E^2$-pages for three dimensions}
In this Appendix, we summarize the $E^2$-pages ($E^2_{p,n}$) for three dimensions that we have used in the main text. 

For TRS-breaking spinful insulators with non-magnetic (32 types) point group symmetry, the $E^1$-page has the 2-fold Bott periodicity $E^1_{p,n+2} = E^1_{p,n}$. 
From (\ref{eq:e1_pt}), $E^1_{p,n}$ is the direct sum of free Abelian groups at $p$-cells in the symmetry class with the grading $n$. 
Under the stable equivalence, there is no representation in class AIII insulators, meaning that $E^1_{p,1}$ vanishes for all $p$ and so is $E^2_{p,1}$. 
Taking the homology of the first differential (\ref{eq:d1_map}), we have the $E^2$-page $E^2_{p,0}$, which is listed in Table \ref{tab:e2_nonmag} in the following form 
\begin{align}
    \left[ E^2_{0,0},E^2_{1,0},E^2_{2,0},E^2_{3,0}\right]. 
\end{align}
As you can see, the second and third differentials $d^r_{p,n}: E^r_{p,n} \to E^r_{p-r,n+r-1}, r=2,3$, are trivial. 
Therefore, the $E^2$-page is the limiting page. 

For TRS-breaking spinful insulators with magnetic (58 types) point group symmetry, the $E^1$-page obeys the 8-fold Bott periodicity $E^1_{p,n+8} = E^1_{p,n}$. 
In the $E^{\infty}$-page, $E^{\infty}_{0,0}, E^{\infty}_{1,-1}, E^{\infty}_{2,-2}$ and $E^{\infty}_{3,-3}$ constitute the $K$-homology $K^G_0(\mathbb{E}^3)$ that classifies the three dimensional topological insulators with the symmetry group $G$. 
In Table \ref{tab:e2_mag}, we provide the parts of $E^2$-pages with $n \in \{0,-1,-2,-3\}$, which is shown in the following form 
\begin{align}
    \left[
    \begin{array}{llll}
E^2_{0,0} & E^2_{1,0} & E^2_{2,0} & E^2_{3,0} \\
E^2_{0,7} & E^2_{1,-1} & E^2_{2,-2} & E^2_{3,-3} \\
E^2_{0,6} & E^2_{1,-1} & E^2_{2,-2} & E^2_{3,-3} \\
E^2_{0,5} & E^2_{1,-1} & E^2_{2,-2} & E^2_{3,-3} \\
    \end{array}
    \right], 
\end{align}
for 58 magnetic point groups. 
We find that, for $E^2$-pages relevant to the $K$-homology $K^G_0(\mathbb{E}^3)$, the second and third differentials vanish.

\begin{table}
\caption{$E^2$-pages for TRS-breaking spinful topological insulators with non-magnetic point group (PG) symmetry.}
\label{tab:e2_nonmag}
$$
\begin{array}{cc}
\mbox{PG} & \left[ E^2_{0,0}, E^2_{1,0}, E^2_{2,0}, E^2_{3,0} \right]  \\
\hline
 C_1 & \left[
\begin{array}{cccc}
 0 & 0 & 0 & \Z \\
\end{array}
\right] \\
 C_i & \left[
\begin{array}{cccc}
 \Z & 0 & \Z_2 & 0 \\
\end{array}
\right] \\
 C_2 & \left[
\begin{array}{cccc}
 0 & \Z & 0 & \Z \\
\end{array}
\right] \\
 C_s & \left[
\begin{array}{cccc}
 0 & 0 & \Z & 0 \\
\end{array}
\right] \\
 C_{2h} & \left[
\begin{array}{cccc}
 \Z & 0 & \Z & 0 \\
\end{array}
\right] \\
 D_2 & \left[
\begin{array}{cccc}
 0 & \Z^3 & 0 & \Z \\
\end{array}
\right] \\
 C_{2v} & \left[
\begin{array}{cccc}
 0 & 0 & \Z^2 & 0 \\
\end{array}
\right] \\
 D_{2h} & \left[
\begin{array}{cccc}
 \Z & 0 & \Z^3 & 0 \\
\end{array}
\right] \\
 C_4 & \left[
\begin{array}{cccc}
 0 & \Z^3 & 0 & \Z \\
\end{array}
\right] \\
 S_4 & \left[
\begin{array}{cccc}
 \Z^2 & 0 & \Z_2 & 0 \\
\end{array}
\right] \\
 C_{4h} & \left[
\begin{array}{cccc}
 \Z^3 & 0 & \Z & 0 \\
\end{array}
\right] \\
 D_4 & \left[
\begin{array}{cccc}
 0 & \Z^4 & 0 & \Z \\
\end{array}
\right] \\
 C_{4v} & \left[
\begin{array}{cccc}
 0 & \Z & \Z^2 & 0 \\
\end{array}
\right] \\
 D_{2d} & \left[
\begin{array}{cccc}
 \Z & \Z & \Z & 0 \\
\end{array}
\right] \\
 D_{4h} & \left[
\begin{array}{cccc}
 \Z^2 & 0 & \Z^3 & 0 \\
\end{array}
\right] \\
 C_3 & \left[
\begin{array}{cccc}
 0 & \Z^2 & 0 & \Z \\
\end{array}
\right] \\
 C_{3i} & \left[
\begin{array}{cccc}
 \Z^3 & 0 & \Z_2 & 0 \\
\end{array}
\right] \\
 D_3 & \left[
\begin{array}{cccc}
 0 & \Z^2 & 0 & \Z \\
\end{array}
\right] \\
 C_{3v} & \left[
\begin{array}{cccc}
 0 & \Z & \Z & 0 \\
\end{array}
\right] \\
 D_{3d} & \left[
\begin{array}{cccc}
 \Z^2 & 0 & \Z & 0 \\
\end{array}
\right] \\
 C_6 & \left[
\begin{array}{cccc}
 0 & \Z^5 & 0 & \Z \\
\end{array}
\right] \\
 C_{3h} & \left[
\begin{array}{cccc}
 \Z^2 & 0 & \Z & 0 \\
\end{array}
\right] \\
 C_{6h} & \left[
\begin{array}{cccc}
 \Z^5 & 0 & \Z & 0 \\
\end{array}
\right] \\
 D_6 & \left[
\begin{array}{cccc}
 0 & \Z^5 & 0 & \Z \\
\end{array}
\right] \\
 C_{6v} & \left[
\begin{array}{cccc}
 0 & \Z^2 & \Z^2 & 0 \\
\end{array}
\right] \\
 D_{3h} & \left[
\begin{array}{cccc}
 \Z & 0 & \Z^2 & 0 \\
\end{array}
\right] \\
 D_{6h} & \left[
\begin{array}{cccc}
 \Z^3 & 0 & \Z^3 & 0 \\
\end{array}
\right] \\
 T & \left[
\begin{array}{cccc}
 0 & \Z^3 & 0 & \Z \\
\end{array}
\right] \\
 T_h & \left[
\begin{array}{cccc}
 \Z^3 & 0 & \Z & 0 \\
\end{array}
\right] \\
 O & \left[
\begin{array}{cccc}
 0 & \Z^4 & 0 & \Z \\
\end{array}
\right] \\
 T_d & \left[
\begin{array}{cccc}
 \Z & \Z & \Z & 0 \\
\end{array}
\right] \\
 O_h & \left[
\begin{array}{cccc}
 \Z^3 & 0 & \Z^2 & 0 \\
\end{array}
\right] \\
\end{array}
$$
\end{table}

\begin{widetext}

\begin{table}
\caption{$E^2$-pages for TRS-breaking spinful topological insulators with magnetic point group (MPG) symmetry.}
\label{tab:e2_mag}
\begin{align*}
\begin{array}{cc}
\mbox{MPG} & \left[
    \begin{array}{llll}
E^2_{0,0} & E^2_{1,0} & E^2_{2,0} & E^2_{3,0} \\
E^2_{0,-1} & E^2_{1,-1} & E^2_{2,-1} & E^2_{3,-1} \\
E^2_{0,-2} & E^2_{1,-2} & E^2_{2,-2} & E^2_{3,-2} \\
E^2_{0,-3} & E^2_{1,-3} & E^2_{2,-3} & E^2_{3,-3} \\
    \end{array}
    \right] \\
\hline
 \bar 1' & \left[
\begin{array}{cccc}
 0 & 0 & \Z_2 & 0 \\
 0 & 0 & 0 & 0 \\
 0 & 0 & 0 & \Z \\
 \Z_2 & 0 & 0 & 0 \\
\end{array}
\right] \\
 2'& \left[
\begin{array}{cccc}
 0 & \Z_2 & 0 & \Z \\
 0 & 0 & 0 & 0 \\
 0 & 0 & \Z_2 & 0 \\
 0 & 0 & 0 & 0 \\
\end{array}
\right] \\
 m' & \left[
\begin{array}{cccc}
 0 & 0 & \Z_2 & 0 \\
 0 & 0 & 0 & 0 \\
 0 & 0 & 0 & \Z \\
 0 & 0 & 0 & 0 \\
\end{array}
\right] \\
 2'/m & \left[
\begin{array}{cccc}
 0 & \Z_2 & 0 & 0 \\
 0 & 0 & 0 & 0 \\
 0 & 0 & \Z & 0 \\
 0 & 0 & 0 & 0 \\
\end{array}
\right] \\
 2/m' & \left[
\begin{array}{cccc}
 0 & \Z & \Z_2 & 0 \\
 0 & 0 & 0 & 0 \\
 0 & 0 & 0 & \Z \\
 0 & 0 & 0 & 0 \\
\end{array}
\right] \\
 2'/m' & \left[
\begin{array}{cccc}
 \Z & \Z_2 & \Z_2 & 0 \\
 0 & 0 & 0 & 0 \\
 0 & 0 & \Z_2 & 0 \\
 0 & 0 & 0 & 0 \\
\end{array}
\right] \\
 2'2'2 & \left[
\begin{array}{cccc}
 \Z_2 & \Z_2 & 0 & \Z \\
 0 & 0 & 0 & 0 \\
 0 & \Z & \Z_2 & 0 \\
 0 & 0 & 0 & 0 \\
\end{array}
\right] \\
 m'm2' & \left[
\begin{array}{cccc}
 0 & \Z_2 & 0 & 0 \\
 0 & 0 & 0 & 0 \\
 0 & 0 & \Z & 0 \\
 0 & 0 & 0 & 0 \\
\end{array}
\right] \\
 m'm'2 & \left[
\begin{array}{cccc}
 0 & \Z & \Z_2 & 0 \\
 0 & 0 & 0 & 0 \\
 0 & 0 & 0 & \Z \\
 0 & 0 & 0 & 0 \\
\end{array}
\right] \\
 m'mm & \left[
\begin{array}{cccc}
 0 & \Z_2^2 & 0 & 0 \\
 0 & 0 & 0 & 0 \\
 0 & 0 & \Z^2 & 0 \\
 0 & 0 & 0 & 0 \\
\end{array}
\right] \\
 m'm'm & \left[
\begin{array}{cccc}
 \Z & \Z_2 & 0 & 0 \\
 0 & 0 & 0 & 0 \\
 0 & 0 & \Z & 0 \\
 0 & 0 & 0 & 0 \\
\end{array}
\right] \\
 m'm'm' & \left[
\begin{array}{cccc}
 0 & \Z^3 & \Z_2 & 0 \\
 0 & 0 & 0 & 0 \\
 0 & 0 & 0 & \Z \\
 0 & 0 & 0 & 0 \\
\end{array}
\right] \\
 4' & \left[
\begin{array}{cccc}
 0 & \Z_2 & 0 & \Z \\
 0 & 0 & 0 & 0 \\
 0 & \Z & \Z_2 & 0 \\
 0 & 0 & 0 & 0 \\
\end{array}
\right] \\
\end{array}
&&
\begin{array}{cc}
 \bar{4}' & \left[
\begin{array}{cccc}
 0 & \Z & \Z_2 & 0 \\
 0 & 0 & 0 & 0 \\
 0 & \Z_2 & 0 & \Z \\
 0 & 0 & 0 & 0 \\
\end{array}
\right] \\
 4'/m & \left[
\begin{array}{cccc}
 \Z & \Z_2 & 0 & 0 \\
 0 & 0 & 0 & 0 \\
 0 & 0 & \Z & 0 \\
 0 & 0 & 0 & 0 \\
\end{array}
\right] \\
 4/m' & \left[
\begin{array}{cccc}
 0 & \Z^2 & \Z_2 & 0 \\
 0 & 0 & 0 & 0 \\
 0 & \Z & 0 & \Z \\
 0 & 0 & 0 & 0 \\
\end{array}
\right] \\
 4'/m' & \left[
\begin{array}{cccc}
 \Z & \Z_2 & \Z_2 & 0 \\
 0 & 0 & 0 & 0 \\
 \Z & 0 & \Z_2 & 0 \\
 0 & 0 & 0 & 0 \\
\end{array}
\right] \\
 4'22' & \left[
\begin{array}{cccc}
 \Z_2 & \Z\oplus\Z_2 & 0 & \Z \\
 0 & 0 & 0 & 0 \\
 0 & \Z^2 & \Z_2 & 0 \\
 0 & 0 & 0 & 0 \\
\end{array}
\right] \\
 42'2' & \left[
\begin{array}{cccc}
 \Z_2^3 & \Z_2 & 0 & \Z \\
 0 & 0 & 0 & 0 \\
 0 & \Z^3 & \Z_2 & 0 \\
 0 & 0 & 0 & 0 \\
\end{array}
\right] \\
 4'm'm & \left[
\begin{array}{cccc}
 0 & \Z_2 & \Z & 0 \\
 0 & 0 & 0 & 0 \\
 0 & 0 & \Z & 0 \\
 0 & 0 & 0 & 0 \\
\end{array}
\right] \\
 4m'm' & \left[
\begin{array}{cccc}
 0 & \Z^3 & \Z_2 & 0 \\
 0 & 0 & 0 & 0 \\
 0 & 0 & 0 & \Z \\
 0 & 0 & 0 & 0 \\
\end{array}
\right] \\
 \bar{4}'2'm & \left[
\begin{array}{cccc}
 \Z_2 & \Z_2 & \Z & 0 \\
 0 & 0 & 0 & 0 \\
 0 & 0 & \Z & 0 \\
 0 & 0 & 0 & 0 \\
\end{array}
\right] \\
 \bar{4}'2m' & \left[
\begin{array}{cccc}
 0 & \Z^2 & \Z_2 & 0 \\
 0 & 0 & 0 & 0 \\
 0 & \Z & 0 & \Z \\
 0 & 0 & 0 & 0 \\
\end{array}
\right] \\
 \bar{4}2'm' & \left[
\begin{array}{cccc}
 \Z^2 & \Z_2 & \Z_2 & 0 \\
 0 & 0 & 0 & 0 \\
 0 & 0 & \Z_2 & 0 \\
 0 & 0 & 0 & 0 \\
\end{array}
\right] \\
 4/m'mm & \left[
\begin{array}{cccc}
 \Z_2 & \Z_2^2 & 0 & 0 \\
 0 & 0 & 0 & 0 \\
 0 & \Z & \Z^2 & 0 \\
 0 & 0 & 0 & 0 \\
\end{array}
\right] \\
 4'/mm'm & \left[
\begin{array}{cccc}
 \Z & \Z_2 & \Z & 0 \\
 0 & 0 & 0 & 0 \\
 0 & 0 & \Z^2 & 0 \\
 0 & 0 & 0 & 0 \\
\end{array}
\right] \\
 4'/m'm'm & \left[
\begin{array}{cccc}
 \Z & \Z\oplus\Z_2 & 0 & 0 \\
 0 & 0 & 0 & 0 \\
 0 & 0 & \Z & 0 \\
 0 & 0 & 0 & 0 \\
\end{array}
\right] \\
\end{array}
&&
\begin{array}{cc}
 4/mm'm' & \left[
\begin{array}{cccc}
 \Z^3 & \Z_2 & 0 & 0 \\
 0 & 0 & 0 & 0 \\
 0 & 0 & \Z & 0 \\
 0 & 0 & 0 & 0 \\
\end{array}
\right] \\
 4/m'm'm' & \left[
\begin{array}{cccc}
 0 & \Z^4 & \Z_2 & 0 \\
 0 & 0 & 0 & 0 \\
 0 & 0 & 0 & \Z \\
 0 & 0 & 0 & 0 \\
\end{array}
\right] \\
\bar 3' & \left[
\begin{array}{cccc}
 0 & \Z & \Z_2 & 0 \\
 0 & 0 & 0 & 0 \\
 0 & \Z & 0 & \Z \\
 \Z_2 & 0 & 0 & 0 \\
\end{array}
\right] \\
 32' & \left[
\begin{array}{cccc}
 \Z_2^2 & \Z_2 & 0 & \Z \\
 0 & 0 & 0 & 0 \\
 0 & \Z^2 & \Z_2 & 0 \\
 0 & 0 & 0 & 0 \\
\end{array}
\right] \\
 3m' & \left[
\begin{array}{cccc}
 0 & \Z^2 & \Z_2 & 0 \\
 0 & 0 & 0 & 0 \\
 0 & 0 & 0 & \Z \\
 0 & 0 & 0 & 0 \\
\end{array}
\right] \\
 \bar{3}'m & \left[
\begin{array}{cccc}
 \Z_2 & \Z_2 & 0 & 0 \\
 0 & 0 & 0 & 0 \\
 0 & \Z & \Z & 0 \\
 0 & 0 & 0 & 0 \\
\end{array}
\right] \\
 \bar{3}'m' & \left[
\begin{array}{cccc}
 0 & \Z^2 & \Z_2 & 0 \\
 0 & 0 & 0 & 0 \\
 0 & 0 & 0 & \Z \\
 0 & 0 & 0 & 0 \\
\end{array}
\right] \\
 \bar{3}m' & \left[
\begin{array}{cccc}
 \Z^3 & \Z_2 & \Z_2 & 0 \\
 0 & 0 & 0 & 0 \\
 0 & 0 & 0 & 0 \\
 0 & 0 & 0 & 0 \\
\end{array}
\right] \\
 6' & \left[
\begin{array}{cccc}
 0 & \Z\oplus\Z_2 & 0 & \Z \\
 0 & 0 & 0 & 0 \\
 0 & \Z & \Z_2 & 0 \\
 0 & 0 & 0 & 0 \\
\end{array}
\right] \\
 \bar{6}' & \left[
\begin{array}{cccc}
 0 & \Z & \Z_2 & 0 \\
 0 & 0 & 0 & 0 \\
 0 & \Z & 0 & \Z \\
 0 & 0 & 0 & 0 \\
\end{array}
\right] \\
 6'/m & \left[
\begin{array}{cccc}
 \Z & \Z_2 & 0 & 0 \\
 0 & 0 & 0 & 0 \\
 \Z & 0 & \Z & 0 \\
 0 & 0 & 0 & 0 \\
\end{array}
\right] \\
 6/m' & \left[
\begin{array}{cccc}
 0 & \Z^3 & \Z_2 & 0 \\
 0 & 0 & 0 & 0 \\
 0 & \Z^2 & 0 & \Z \\
 0 & 0 & 0 & 0 \\
\end{array}
\right] \\
6'/m' & \left[
\begin{array}{cccc}
 \Z^2 & \Z_2 & \Z_2 & 0 \\
 0 & 0 & 0 & 0 \\
 \Z & 0 & \Z_2 & 0 \\
 0 & 0 & 0 & 0 \\
\end{array}
\right] \\
 6'22' & \left[
\begin{array}{cccc}
 \Z_2^2 & \Z_2 & 0 & \Z \\
 0 & 0 & 0 & 0 \\
 0 & \Z^2 & \Z_2 & 0 \\
 0 & 0 & 0 & 0 \\
\end{array}
\right] \\
\end{array}
\end{align*}
\end{table}

\begin{table}[]
    \caption*{(Continued from the previous page)}
\begin{align*}
\begin{array}{cc}
 62'2' & \left[
\begin{array}{cccc}
 \Z_2^5 & \Z_2 & 0 & \Z \\
 0 & 0 & 0 & 0 \\
 0 & \Z^5 & \Z_2 & 0 \\
 0 & 0 & 0 & 0 \\
\end{array}
\right] \\
 6'mm' & \left[
\begin{array}{cccc}
 0 & \Z\oplus\Z_2 & 0 & 0 \\
 0 & 0 & 0 & 0 \\
 0 & 0 & \Z & 0 \\
 0 & 0 & 0 & 0 \\
\end{array}
\right] \\
 6m'm' & \left[
\begin{array}{cccc}
 0 & \Z^5 & \Z_2 & 0 \\
 0 & 0 & 0 & 0 \\
 0 & 0 & 0 & \Z \\
 0 & 0 & 0 & 0 \\
\end{array}
\right] \\
 \bar{6}'m'2 & \left[
\begin{array}{cccc}
 0 & \Z^2 & \Z_2 & 0 \\
 0 & 0 & 0 & 0 \\
 0 & 0 & 0 & \Z \\
 0 & 0 & 0 & 0 \\
\end{array}
\right] \\
 \bar{6}'m2' & \left[
\begin{array}{cccc}
 \Z_2 & \Z_2 & 0 & 0 \\
 0 & 0 & 0 & 0 \\
 0 & \Z & \Z & 0 \\
 0 & 0 & 0 & 0 \\
\end{array}
\right] \\
 \bar{6}m'2' & \left[
\begin{array}{cccc}
 \Z^2 & \Z_2 & 0 & 0 \\
 0 & 0 & 0 & 0 \\
 0 & 0 & \Z & 0 \\
 0 & 0 & 0 & 0 \\
\end{array}
\right] \\
 6/m'mm & \left[
\begin{array}{cccc}
 \Z_2^2 & \Z_2^2 & 0 & 0 \\
 0 & 0 & 0 & 0 \\
 0 & \Z^2 & \Z^2 & 0 \\
 0 & 0 & 0 & 0 \\
\end{array}
\right] \\
 6'/mmm' & \left[
\begin{array}{cccc}
 \Z & \Z_2^2 & 0 & 0 \\
 0 & 0 & 0 & 0 \\
 0 & 0 & \Z^2 & 0 \\
 0 & 0 & 0 & 0 \\
\end{array}
\right] \\
 6'/m'mm' & \left[
\begin{array}{cccc}
 \Z^2 & \Z_2 & 0 & 0 \\
 0 & 0 & 0 & 0 \\
 0 & 0 & \Z & 0 \\
 0 & 0 & 0 & 0 \\
\end{array}
\right] \\
\end{array}
&&
\begin{array}{cc}
 6/mm'm' & \left[
\begin{array}{cccc}
 \Z^5 & \Z_2 & 0 & 0 \\
 0 & 0 & 0 & 0 \\
 0 & 0 & \Z & 0 \\
 0 & 0 & 0 & 0 \\
\end{array}
\right] \\
 6/m'm'm' & \left[
\begin{array}{cccc}
 0 & \Z^5 & \Z_2 & 0 \\
 0 & 0 & 0 & 0 \\
 0 & 0 & 0 & \Z \\
 0 & 0 & 0 & 0 \\
\end{array}
\right] \\
 m'\bar{3}' & \left[
\begin{array}{cccc}
 0 & \Z^2 & \Z_2 & 0 \\
 0 & 0 & 0 & 0 \\
 0 & \Z & 0 & \Z \\
 0 & 0 & 0 & 0 \\
\end{array}
\right] \\
 4'32' & \left[
\begin{array}{cccc}
 \Z_2^3 & \Z_2 & 0 & \Z \\
 0 & 0 & 0 & 0 \\
 0 & \Z^3 & \Z_2 & 0 \\
 0 & 0 & 0 & 0 \\
\end{array}
\right] \\
 \bar{4}'3m' & \left[
\begin{array}{cccc}
 0 & \Z^3 & \Z_2 & 0 \\
 0 & 0 & 0 & 0 \\
 0 & 0 & 0 & \Z \\
 0 & 0 & 0 & 0 \\
\end{array}
\right] \\
 m'\bar{3}'m & \left[
\begin{array}{cccc}
 \Z\oplus\Z_2 & \Z_2 & 0 & 0 \\
 0 & 0 & 0 & 0 \\
 0 & \Z & \Z & 0 \\
 0 & 0 & 0 & 0 \\
\end{array}
\right] \\
 m\bar{3}m' & \left[
\begin{array}{cccc}
 \Z^3 & \Z_2 & 0 & 0 \\
 0 & 0 & 0 & 0 \\
 0 & 0 & \Z & 0 \\
 0 & 0 & 0 & 0 \\
\end{array}
\right] \\
 m'\bar{3}'m' & \left[
\begin{array}{cccc}
 0 & \Z^4 & \Z_2 & 0 \\
 0 & 0 & 0 & 0 \\
 0 & 0 & 0 & \Z \\
 0 & 0 & 0 & 0 \\
\end{array}
\right] \\
\end{array}
\end{align*}
\end{table}

\end{widetext}


\begin{thebibliography}{9}
\bibitem{note1}
In the rest of this paper, the notation $K^G_n(\mathbb{E}^3)$ means the abbreviation of the relative $K$-group $K^G_n(D^3,\partial D^3)$ with $D^3 (\cong \mathbb{E}^3)$ the three dimensional disc and $\partial D^3 \cong S^2$ its boundary.
\bibitem{skeleton}
Note that lower-dimensional cells and skeletons can be an empty set as in Fig. $\ref{fig5}$, where zero-dimensional cells and skeletons are absent.
\bibitem{adiabaticpump}
We can also interpret $E_{p+1, n}^1$ as the $(p+n)$-th graded ``adiabatic pump" on $(p+1)$-cells\cite{shiozaki1}. In this interpretation, Im $d_{p+1,n}^1$ can be regarded as the $(p+n)$-th SPT phase on $p$-cells that is pumped from $(p+1)$-cells. Also, $E_{p,n}^2$ is the $(p+n)$-th graded SPT on $p$-cells that creates no boundary states on the adjacent $(p-1)$-cells and cannot be created from the adjacent $(p+1)$-cells by adiabatic pump.
\bibitem{theoremb}
Another explanation is as follows: If $H$ has an additional uniform mass term $\bm{M_0}$, then $H$ is topologically trivial\cite{chiu} in terms of $E_{0,0}^\infty$, which implies the theorem B.
\end{thebibliography}
\end{document}